\documentclass[final,3p,times,twocolumn]{elsarticle}
\usepackage{graphicx}
\usepackage{dcolumn}
\usepackage{amssymb,amsmath,amsthm,mathrsfs}
\usepackage{epstopdf}
\usepackage{lipsum}
\usepackage{hyperref}
\usepackage{empheq}
\usepackage{color}
\usepackage[normalem]{ulem} 

\newcommand{\be}{\begin{equation}}
\newcommand{\ee}{\end{equation}}
 \newcommand{\bea}{\begin{eqnarray}}
\newcommand{\eea}{\end{eqnarray}}

\newcommand{\CLns}{{\tt ${\mathcal C}$osmo${\mathcal L}$attice}}

\begin{document}

\title{Energy distribution and equation of state of the early Universe:\\matching the end of inflation and the onset of radiation domination}

\author[label1]{Stefan Antusch}
\author[label2]{Daniel G. Figueroa}
\author[label1]{Kenneth Marschall}
\author[label1]{Francisco Torrenti}

\address[label1]{Department of Physics, University of Basel, Klingelbergstr. 82, CH-4056 Basel, Switzerland.}
\address[label2]{Instituto de Física Corpuscular (IFIC), CSIC-Universitat de Valencia, Spain.}

\begin{abstract}
We study the energy distribution and equation of state of the universe between the end of inflation and the onset of radiation domination (RD), considering observationally consistent single-field inflationary scenarios, with a potential 'flattening' at large field values, and a monomial shape $V(\phi) \propto |\phi|^p$ around the origin. As a proxy for (p)reheating, we include a quadratic interaction $g^2\phi^2X^2$ between the inflaton $\phi$ and a light scalar `daughter' field $X$, with $g^2>0$. We capture the non-perturbative and non-linear nature of the system dynamics with lattice simulations, obtaining that: $i)$ the final energy transferred to $X$ depends only on $p$, not on $g^2$, ; $ii)$ the final transfer of energy is always negligible for $2 \leq p < 4$, and of order $\sim 50\%$ for $p \geq 4$; $iii)$ the system goes at late times to matter-domination for $p = 2$, and always to RD for $p > 2$. In the latter case we calculate the number of e-folds until RD, significantly reducing the uncertainty in the inflationary observables $n_s$ and $r$.
\end{abstract}

\maketitle

\textbf{Introduction -.}  Cosmological observations strongly support the idea of an inflationary period in the early universe \cite{Starobinsky:1980te,Guth:1980zm,Linde:1981mu,Albrecht:1982wi}. Inflation must be followed by a \textit{(p)reheating} stage, where most of the energy in the universe is transferred into light particle species, with only one strong requisite: the universe must arrive at a radiation dominated (RD) thermal state before the start of Big Bang Nucleosynthesis (BBN), at temperatures $T_{\rm BBN}\sim 1\text{MeV}$. The state of the universe at BBN is based on the Standard Model (SM) particle content, which is fairly known. However, the way the universe arrives at this state from the previous inflation stage is largely unknown, and depends strongly on the underlying particle physics model. 

Measurements of the cosmic microwave background (CMB) provide an upper bound on the inflationary Hubble rate,  $H_{\rm inf} \lesssim 6.6\times10^{13}$ GeV~\cite{Akrami:2018odb,Ade:2018gkx}, corresponding to energy scales just below $\sim 10^{16}$ GeV. The energy gap between the end of inflation and the onset of BBN may therefore span up to $\sim 19$ orders of magnitude. Characterizing this {\it primordial dark age} period is important, as it represents a natural 'cosmological window' to probe {\it beyond the SM} (BSM) physics, potentially displaying a very rich phenomenology, see~\cite{Bassett:2005xm,Allahverdi:2010xz,Amin:2014eta,Lozanov:2019jxc,Allahverdi:2020bys} for reviews and references therein. Moreover, the {\it equation of state} (EoS) during this period is required for making accurate predictions of inflationary CMB observables, see e.g.~\cite{Dai:2014jja,Martin:2014nya,Munoz:2014eqa,Gong:2015qha}. 

In the context of slow-roll single-field inflation, a {\it preheating} phase emerges when the {\it inflaton} $\phi$, the field responsible for inflation, starts oscillating around the minimum of its potential. In this work we consider a broad class of observationally viable scenarios inspired by $\alpha$-{\it attractors} \cite{Kallosh:2013hoa},
with 'flattening' of the inflaton potential at large field values, and monomial behaviour $V(\phi) \propto |\phi|^p$ around the origin, with $p\geq 2$ including fractional values. The inflaton is directly coupled to a '(p)reheating sector' represented by a light scalar field $X$, which will be called the \textit{daughter field} from now on. We consider a quadratic interaction $g^2\phi^2X^2$, as it does not require the introduction of new mass scales, and serves as a proxy for the leading term in gauge interactions~\cite{Figueroa:2015rqa}. Under these considerations, the universe goes first through a stage of \textit{preheating}, in which the initially homogeneous inflaton condensate fragments via non-perturbative particle production effects, see~\cite{Traschen:1990sw,Kofman:1994rk,Shtanov:1994ce,Khlebnikov:1996mc,Prokopec:1996rr,Kofman:1997yn,Greene:1997fu} for the pioneering studies and e.g.~\cite{Lozanov:2016hid,Figueroa:2016wxr,Lozanov:2017hjm,Giblin:2019nuv} for recent numerical works. Preheating can happen through two separate phenomena: 1) \textit{broad parametric resonance} of the daughter field, in which the inflaton transfers to the former a large amount of its energy exponentially fast, and 2) \textit{self-resonance} of the inflaton, in which the inflaton amplifies its own fluctuations. In both cases, a departure from the (initially homogeneous) inflaton oscillation-averaged EoS is ensued, affecting the following expansion history of the universe.

To investigate (p)reheating we use very long classical lattice simulations in 2+1 dimensions, considering a large range of inflaton-daughter couplings. We are interested in the number of e-folds $\Delta N_{\rm RD}$ from the end of inflation till the onset of RD. Previous works~\cite{Lozanov:2016hid, Lozanov:2017hjm} have obtained this number in the absence of inflaton-daughter interactions, $\Delta N_{\rm RD}|_{g^2=0}$. However, the time scale of the daughter field excitation through broad resonance is faster than via inflaton self-resonance, so $\Delta N_{\rm RD}|_{g^2=0}$ represents only an upper bound to this quantity. Furthermore, recent criticism to (gravitational) reheating in the absence of inflaton couplings to other species~\cite{Figueroa:2018twl,Figueroa:2019paj}, reinforces the idea that the universe is most naturally reheated if inflaton-daughter interactions are present. 

In this Letter we consider a large range of inflaton-daughter interactions, exploring both a large coupling regime (which leads to broad resonance of the daughter field), and a small coupling regime (which recovers the coupling-less results from~\cite{Lozanov:2016hid, Lozanov:2017hjm}). We characterize in detail the energy distribution, EoS, and $\Delta N_{\rm RD}$ as a function of $p$ and $g^2$, and use this information to reduce drastically the uncertainty in the prediction of the inflationary scalar tilt $n_s$ and tensor-to-scalar ratio $r$. Our analysis goes beyond {\it ad-hoc} analytical parametrizations of the post-inflationary EoS~\cite{Saha:2020bis}, and beyond previous numerical works \cite{Podolsky:2005bw,Figueroa:2016wxr,Maity:2018qhi}, which only considered the initial preheating and early non-linear stage for specific choices of $p$. Here we simulate, for the first time, the \textit{very long-term evolution} of the system, for arbitrary values of $p \geq 2$.  \vspace*{-0.3cm}\\

\textbf{Parametric resonance and self-resonance -.} Consider a scenario with an inflaton $\phi$ and daughter field $X$, 
\be \label{eq:inflaton-potential}
V (\phi,X)= \frac{1}{p} \Lambda^4{\rm tanh}^{p} \left( \frac{|\phi|}{M} \right)+\frac{1}{2}g^2\phi^2 X^2\,,
\ee
where $M$ and $\Lambda$ are mass scales, and $g$ is a dimensionless coupling. The first term is the inflaton potential, responsible for slow-roll inflation. The interaction term allows to transfer energy between $\phi$ and $X$.

The inflaton potential features a plateau at $\phi \gg M$ and an inflection point at $\phi_{\rm i} = M {\rm arcsinh} (\sqrt{(p-1)/2})$ $\sim M$. Inflation takes place at field values $\phi \gg \phi_{*} \equiv (M/2) {\rm arcsinh} (\sqrt{2} p m_{\rm pl}/M)$, with $\phi_{*}$ denoting the field value at which $\varepsilon_V(\phi_{*}) \equiv 1$. At $\phi < \phi_{*}$ the inflaton field features an oscillatory regime. For $M/m_{\rm pl} > 1.633$ it holds that $\phi_{\rm i} > \phi_{*}$ $\forall~p \geq 2$, entailing that $\phi$ always oscillates in the positive-curvature region of the potential, which can be approximated around the origin by the power-law $V_{\rm inf}(\phi) \simeq \mu^{4-p} |\phi |^p/p$, with $\mu^{4-p} \equiv \Lambda^4/M^{p}$ for $p \neq 4$, and 
$\mu^{4-p} \equiv \lambda \neq 1$ for $p = 4$. After inflation, the inflaton oscillates initially as a homogeneous condensate, with decaying amplitude $\phi(t) \propto a(t)^{-6/(p+2)}$, and time-dependent oscillation frequency  $\omega^2 = \omega_{*}^2 a^{6(p - 2)/(p+2)}$, with $\omega_{*}^2 \equiv \mu^{4-p} \phi_{*}^{p-2}$. This leads to an EoS \cite{Turner:1983he}
\begin{equation}
    w_{\rm hom} \equiv {\langle p_\phi \rangle_{\rm osc}\over \langle \rho_\phi \rangle_{\rm osc}} = {p-2\over p+2} \ , \label{eq:EoSoscillations-let}
\end{equation}
where $ \langle p_\phi \rangle_{\rm osc}$ and $\langle \rho_\phi \rangle_{\rm osc}$ denote the oscillation-averaged pressure and energy densities of the inflaton. 

For $M \gg m_{\rm pl}$, two preheating effects emerge due to the initial homogeneous oscillations: \textit{parametric resonance} of the daughter field and \textit{self-resonance} of the inflaton. Choosing $a_* = 1$ and re-defining the variables as $dz\equiv a^{-3{(p-2)\over(p+2)}}\omega_{*}dt$, $\varphi \equiv a^{\frac{6}{p+2}} (\phi/ \phi_*)$ and $\chi \equiv a^{\frac{6}{p+2}} (X/ \phi_*)$, the linearized mode equations of the daughter and inflaton fields, during the early oscillatory phase, correspond to oscillator-like equations with time-dependent mass terms $m_{\varphi}^2 \equiv (p-1) |{ \varphi} |^{p-2}$ and $m_{\chi}^2 \equiv q_{\rm res} {\varphi}^2$, where
\be\label{eq:ResParamB}
q_{\rm res} (a) \equiv q_{*} a^{\frac{6 (p-4)}{p+2}} \,, \hspace{0.3cm} q_{*} \equiv g^2\phi_{*}^2 / \omega_{*}^2 \,, \ee
is an effective {\it resonance parameter},
decreasing in time for $p<4$, remaining constant for $p = 4$, and growing for $p>4$. Fluctuations of both fields evolve as $|\delta \chi_k |^2 \propto e^{2\mu_k z}$ and $|\delta \varphi_k |^2 \propto e^{2\nu_k z}$, where  $\mu_k \equiv \mu_k (\kappa,q_{\rm res};p)$ and $\nu_k \equiv \nu_k (\kappa;p)$ are the respective Floquet indices. These functions are positive for some bands of momenta, leading to an exponential growth of the field modes. If parametric resonance is \textit{broad} ($q_{\rm res}\gtrsim 1$), the range of amplified $\delta \chi_k$ is much wider than the one for $\delta \phi_k$. Thus, if both effects are present, the excitation of $\delta \chi_k$ is the dominant one. The daughter field is also excited if the resonance is \textit{narrow} ($q_{\rm res} \lesssim 1$), but this effect is negligible compared to broad resonance. In any case, due to natural limitations of the lattice, it cannot be captured in our simulations. 
On the other hand, the momenta excited during broad resonance scale (modulo scale factor powers) as $p_{\rm br} \sim q_*^{1/4}\omega_* \gtrsim 10^{13}$ GeV, which justifies neglecting a mass term of the daughter field in (\ref{eq:inflaton-potential}) for $m_{X} \ll p_{\rm br}$.   \vspace*{-0.3cm}\\

\textbf{Results -.} We present now our numerical results, obtained from classical lattice simulations of the EOM $\ddot f- a^{-2}\vec{\nabla}^2f+3H\dot{f} = -\partial_f V$ for $f=\{\phi, X\}$ and the Friedmann equation, for different choices of $p$ and $q_{*}$. Details of our lattice formulation are provided in the Appendix. We have performed simulations in 2+1 and 3+1 dimensions, and checked that they are almost identical, see the Appendix for a direct comparison. However, results presented here will be based on simulations in 2+1 dimensions, as they have the advantage of investigating a much larger region of parameter space. We have used a number of lattice sites per dimension ranging from $N=2^7$ to $N=2^{10}$, and explored different infrared and ultraviolet lattice cut-off's, ensuring a range of momenta encompassing well the scales excited by the different resonances. We have simulated the cases $M = 4m_{\rm pl}-10m_{\rm pl}$, which guarantee that the inflaton oscillations occur in the positive-curvature region. \vspace*{-0.3cm}\\

\begin{figure*}
    \begin{center}
    \includegraphics[width=0.31\textwidth]{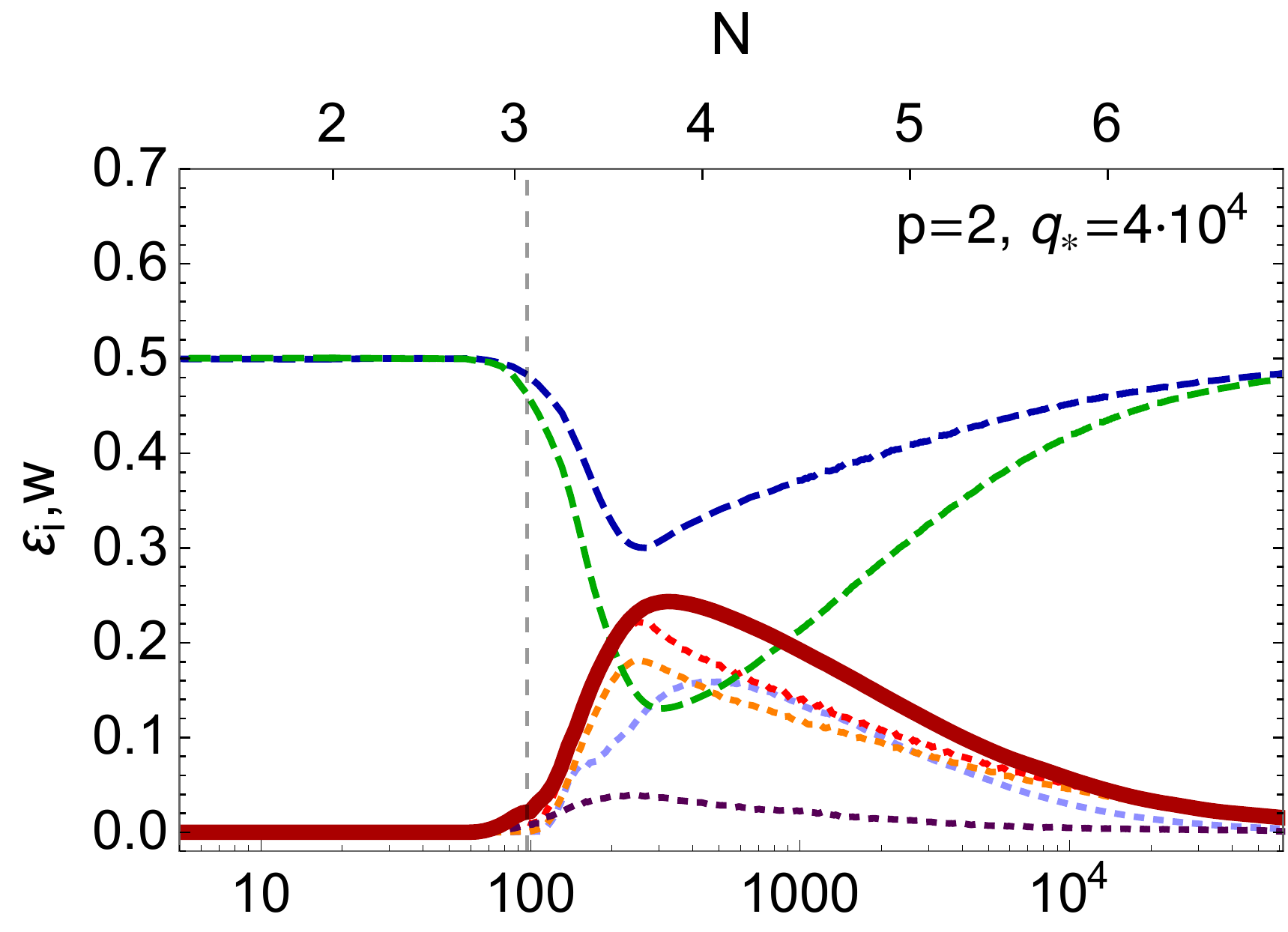} \hspace{0.2cm}
    \includegraphics[width=0.31\textwidth]{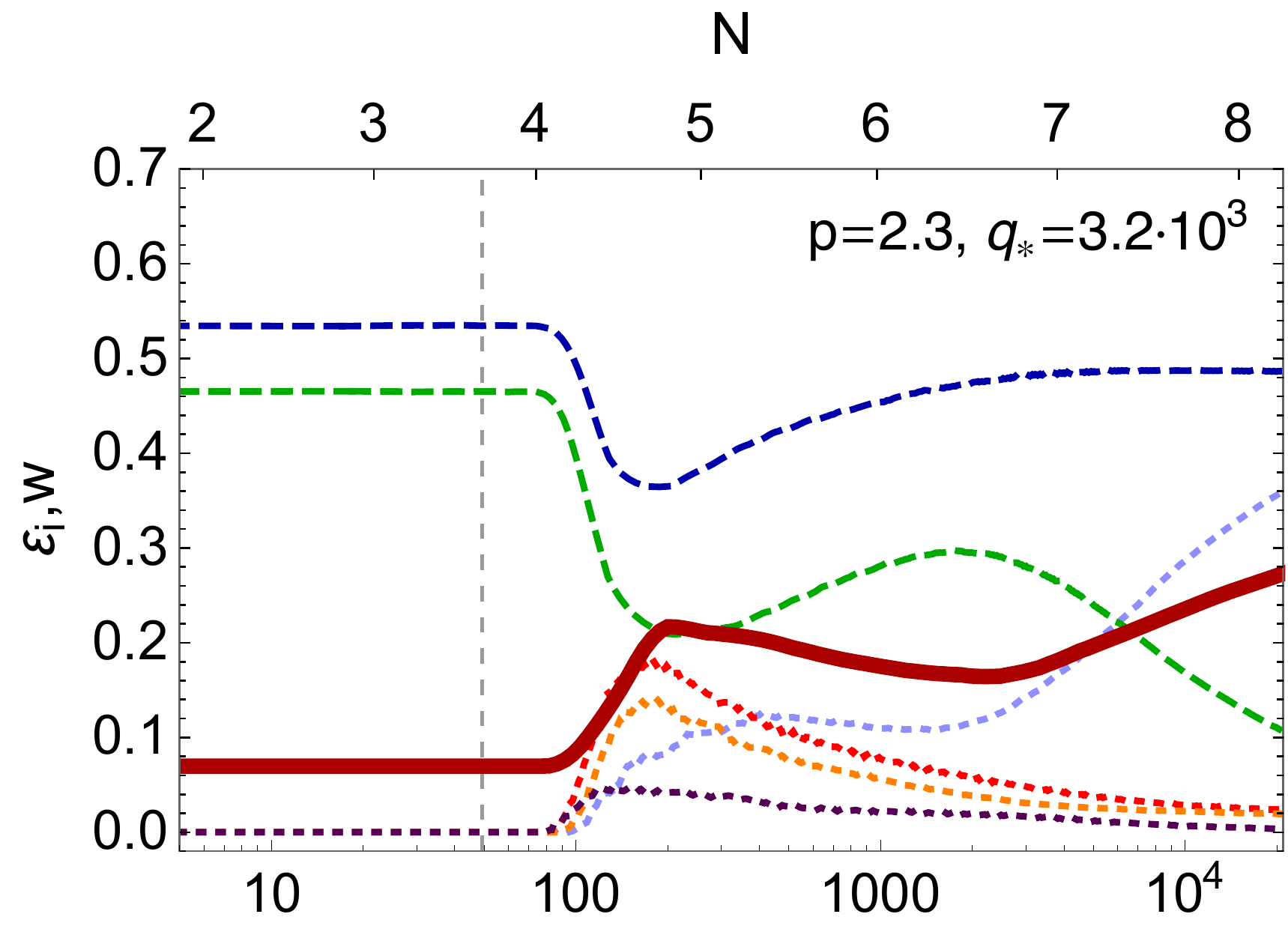}\hspace{0.2cm}
    \includegraphics[width=0.31\textwidth]{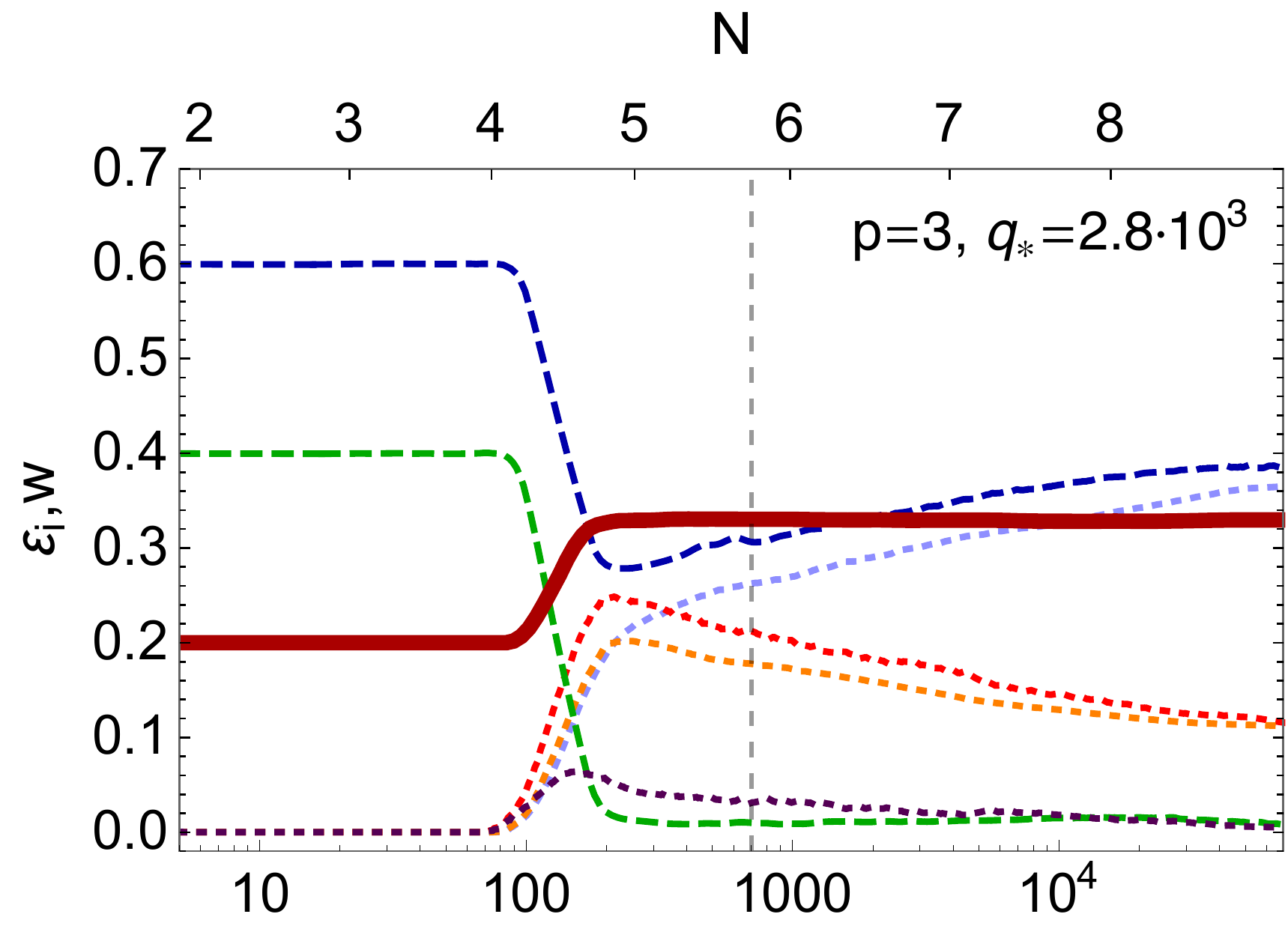} \vspace{0.1cm} \\ 
    \includegraphics[width=0.31\textwidth]{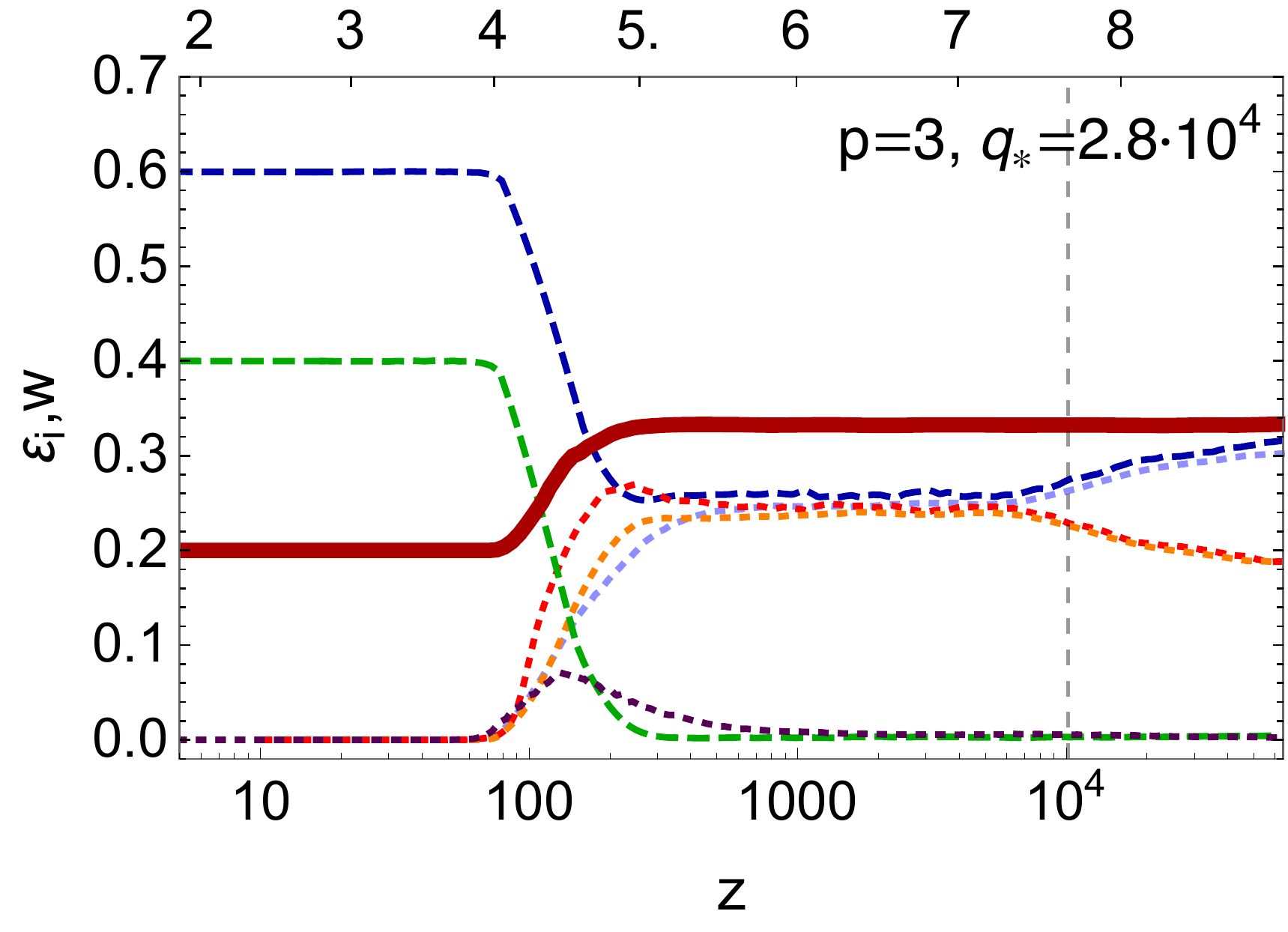} \hspace{0.2cm}
    \includegraphics[width=0.31\textwidth]{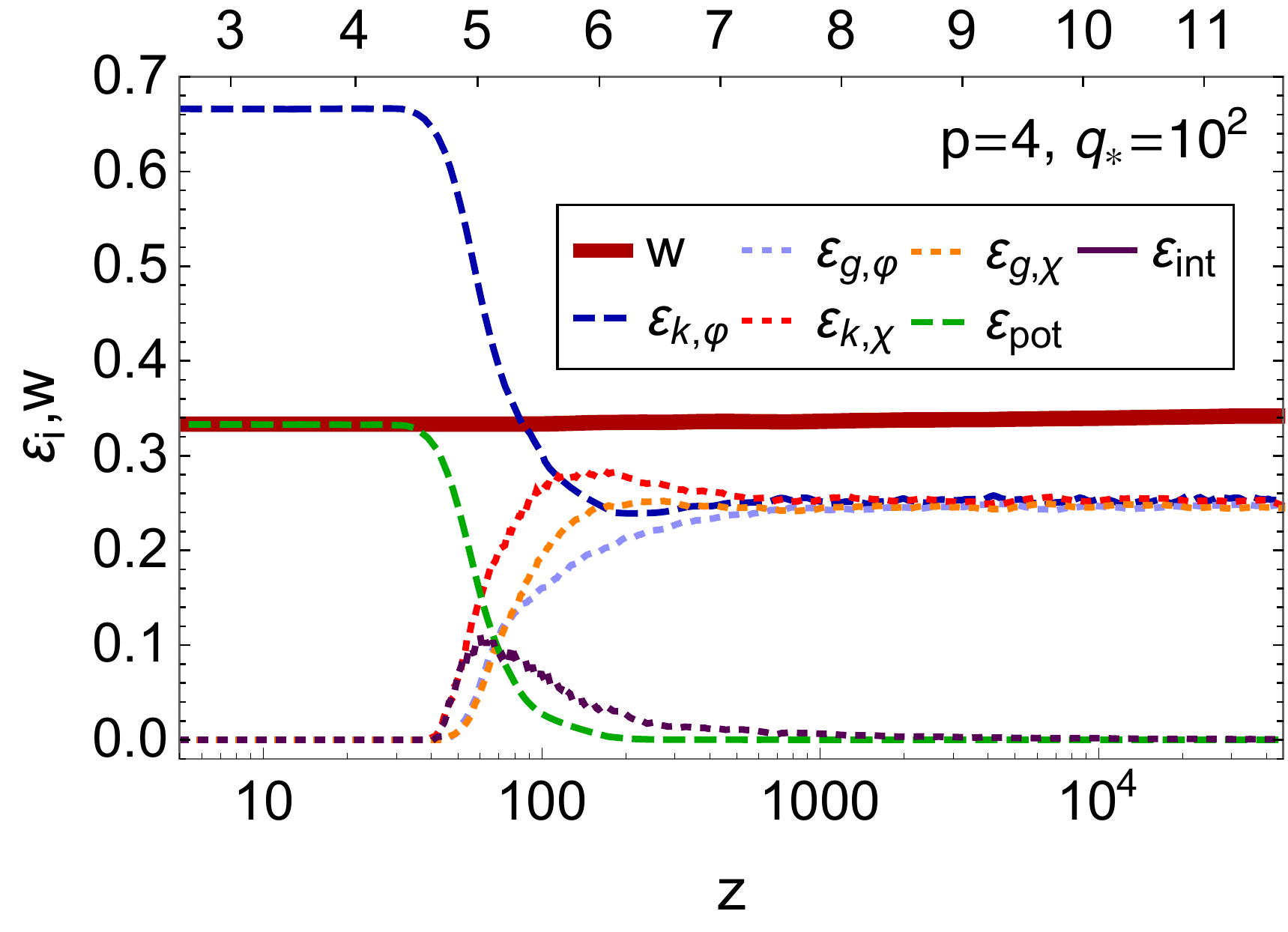} \hspace{0.2cm}
    \includegraphics[width=0.31\textwidth]{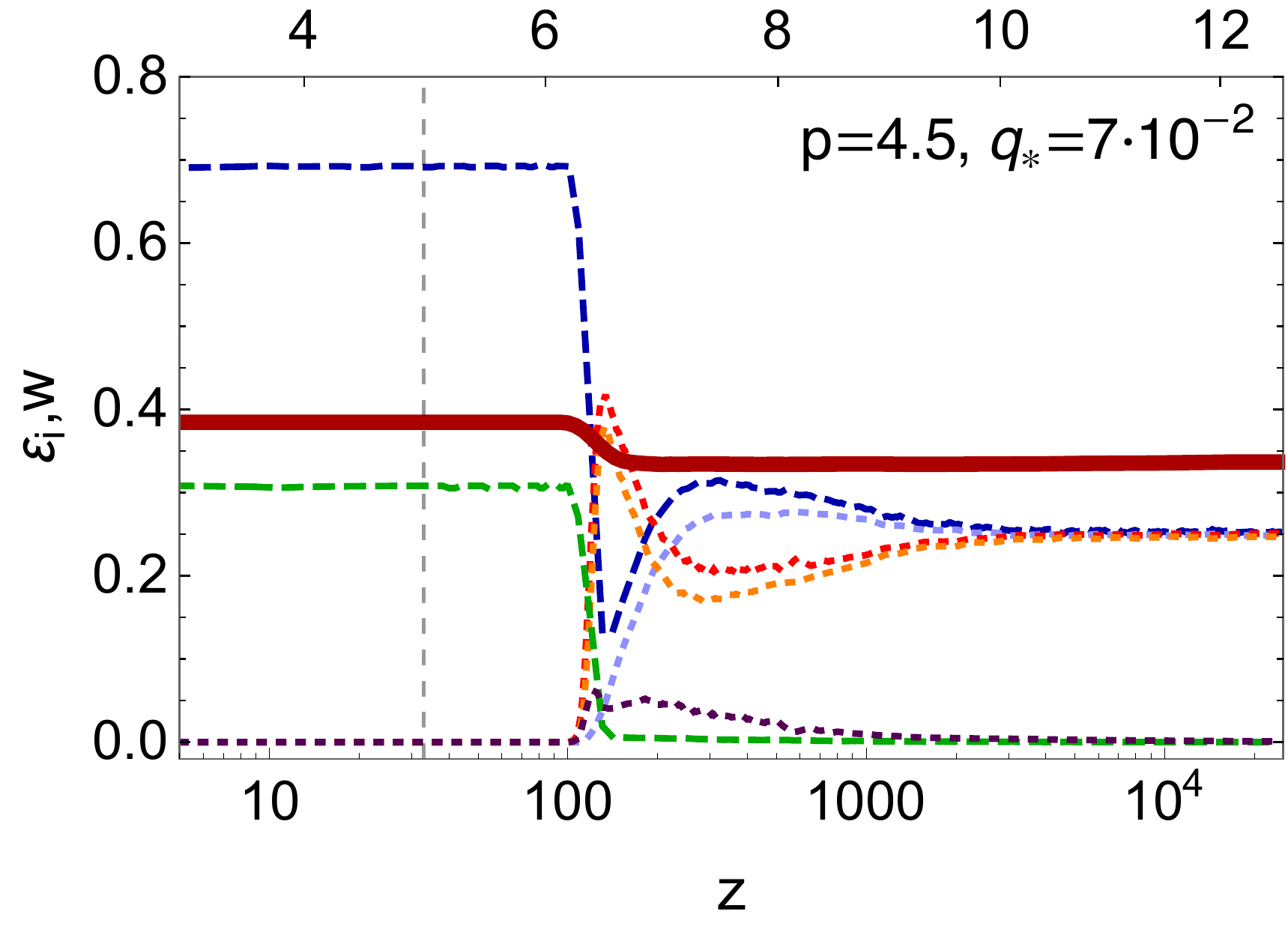} \vspace{-0.3cm}
    \end{center}
    \caption{Evolution of the oscillation-averaged energy ratios and effective equation of state for different choices of $q_*$ and $p$ for $M=10m_\mathrm{pl}$, as a function of time $z$ and number of e-folds $N$. The dashed vertical line in each panel show when $q_{\rm res} = 1$.} \label{fig:EnergyRatios} \vspace*{-0.3cm}
\end{figure*}

\textit{Energy distribution and equation of state}. The different energy density components $\rho_i$ characterize the evolution of the system. Different contributions include kinetic $\rho_{{\rm k},f}$ and gradient $\rho_{{\rm g},f}$ energy components, the inflaton potential $\rho_{\rm pot} = V(\phi)$ [first term in (\ref{eq:inflaton-potential})] and the interaction term $\rho_{\rm int} = {1\over2}g^2\phi^2X^2$. As expected from previous studies, the system {\it virializes} very quickly \cite{Boyanovsky:2003tc,Lozanov:2016hid,Figueroa:2016wxr,Lozanov:2017hjm}, with the fields obeying a relation of the type $\langle {\dot f}^2 \rangle =\langle |\nabla f |^2 \rangle + \left\langle f (\partial V/ \partial f) \right\rangle$, where brackets indicate oscillation and spatial averaging. Introducing energy density ratios as $\varepsilon_i \equiv \rho_i / \sum_{j} \rho_{j}$ (so that $\sum_j \varepsilon_j = 1$ by construction), the virial relations imply 
\bea \langle \varepsilon_{\rm k, \varphi} \rangle & \simeq  & \langle \varepsilon_{\rm g, \varphi}  \rangle + \frac{p}{2} \langle \varepsilon_{\rm pot} \rangle + \langle \varepsilon_{\rm int} \label{eq:Virial1-let}  \rangle \  , \\
\langle \varepsilon_{\rm k, \chi}  \rangle & \simeq & \langle \varepsilon_{\rm g, \chi}  \rangle + \langle \varepsilon_{\rm int}  \rangle \  . \label{eq:Virial2-let} \eea
The instantaneous EoS $w \equiv p / \rho$ is sourced by the different energy contributions as 
\be w  = \varepsilon_{\rm k,\varphi} + \varepsilon_{\rm k,\chi}  - \frac{1}{3} ( \varepsilon_{\rm g,\varphi} + \varepsilon_{\rm g,\chi} ) - ( \varepsilon_{\rm pot} + \varepsilon_{\rm int}) \,. \label{eq:EoS-fEn-let}
\ee
This means that whenever $\varepsilon_{\mathrm{pot}}, \varepsilon_{\rm int}$ become negligible (as it happens e.g.~at later times for $p>2$), then $\varepsilon_{\rm k,\varphi} + \varepsilon_{\rm k,\chi} \simeq 1/2$, which leads to a RD universe with $w = 1/3$. Furthermore, taking averages on both sides of Eq.~(\ref{eq:EoS-fEn-let}) leads to the effective EoS during the first inflaton oscillations: initially $\varepsilon_{\rm k,\varphi} +  \varepsilon_{\rm pot} \simeq 1$ holds, so Eq.~(\ref{eq:Virial1-let}) implies $\langle \varepsilon_{\rm k, \varphi} \rangle  \simeq p/(p+2)$ and $\langle \varepsilon_{\rm pot} \rangle  \simeq 2/(p+2)$, and from there we recover $w_{\rm hom}$ in Eq.~(\ref{eq:EoSoscillations-let}). 

After the initial homogeneous phase, $\lbrace \varepsilon_a \rbrace$ and $w$ evolve very differently depending on the choice of $p$ and $q_{*}$, see panels in Fig.~\ref{fig:EnergyRatios}. We discuss now their evolution, in particular their asymptotic late time behaviour:

$\bullet ~p = 2$\,: There is no self-resonance of the inflaton field, but the daughter field energy grows exponentially fast via broad parametric resonance, as long as $q_{\rm res} > 1$ (top-left panel Fig.~\ref{fig:EnergyRatios}). Thanks to the inflaton-daughter interaction, a growth of the inflaton gradient energy is also induced. However, as $q_{\rm res}$ decreases in time, parametric resonance eventually becomes narrow. For $q_* \gtrsim 6 \cdot 10^3$, broad resonance persists long enough that backreaction effects from the daughter field break the homogeneous inflaton condensate. In such a case, the effective EoS jumps from $w_{\rm hom} \simeq 0$ to a positive value $w_{\rm max} < 1/3$ (closer to $1/3$ the larger $q_{*}$). Gradient energies redshift as $\sim a^{-4}$, whereas the inflaton potential/kinetic energies redshift as $\sim a^{-3}$. Therefore, once $q_{\rm res} < 1$, the daughter energy fractions become gradually negligible, independently of the daughter-inflaton coupling strength. Similarly, the equation of state asymptotically tends to the homogeneous value $w \rightarrow w_{\hom} = 0$. We observe this (otherwise expected) result for the first time, as shorter simulations in previous works were only able to observe a transitory stabilization of the EoS around $w \simeq 0.2$  \cite{Podolsky:2005bw,Maity:2018qhi}. 

$\bullet~2<p<4$\,: The inflaton can now develop fluctuations via self-resonance, but
if $q_{\rm res} \gg 1$, the daughter field energies grow much faster via broad parametric resonance. For $q_{*} \gtrsim 10^{1.9 (4-p)}$, backreaction effects from the daughter field break the initially homogeneous inflaton configuration, making the EoS jump from $w=w_{\rm hom}$ to $w=w_{\rm max} < 1/3$. In fact, a transitory regime of {\it equipartition} can be observed for very large values of $q_*$, equally distributing the energy between the two fields (bottom-left panel of Fig.~\ref{fig:EnergyRatios}). In any case, the resonance eventually becomes narrow when $q_{\rm res} = 1$, and the daughter field energy fractions become gradually negligible. However, the inflaton self-resonance is still present, which triggers a slow cascade of the inflaton spectra towards the ultraviolet, as well as a growth of its gradient energy at the expense of its potential. This phenomenon, originally reported in~\cite{Lozanov:2016hid,Lozanov:2017hjm} in the absence of inflaton-daughter interactions, is observed now remarkably even after the inflaton fragments due to the parametric resonance of the daughter field. As a consequence, the EoS always goes to $w \simeq 1/3$ at sufficiently late times. For $2 < p \lesssim 3$ and certain values of $q_{*}$, the self-resonance is so weak that a temporary decrease of $w$ towards $w_{\rm hom}$ is observed after the end of broad resonance (see top-middle panel), before $w$ goes towards $1/3$ at later times.

$\bullet ~p \geq 4$\,: The resonance parameter $q_{\rm res}$ remains constant for $p=4$, or grows in time for $p>4$. In the latter case, the system always ends up in broad resonance, even if $q_{*} < 1$. As inflaton self-resonance effects are also present, the system never ceases to exchange energy between the two fields at late times. Due to this, it achieves an equilibrium state, in which the energy is evenly distributed: 50\% of the energy is stored in the daughter field, and 50\% in the inflaton.  The equation of state also goes to $w \rightarrow 1/3$ at late times. \vspace{0.2cm}

\begin{figure}
    \centering
    \includegraphics[height=5.1cm]{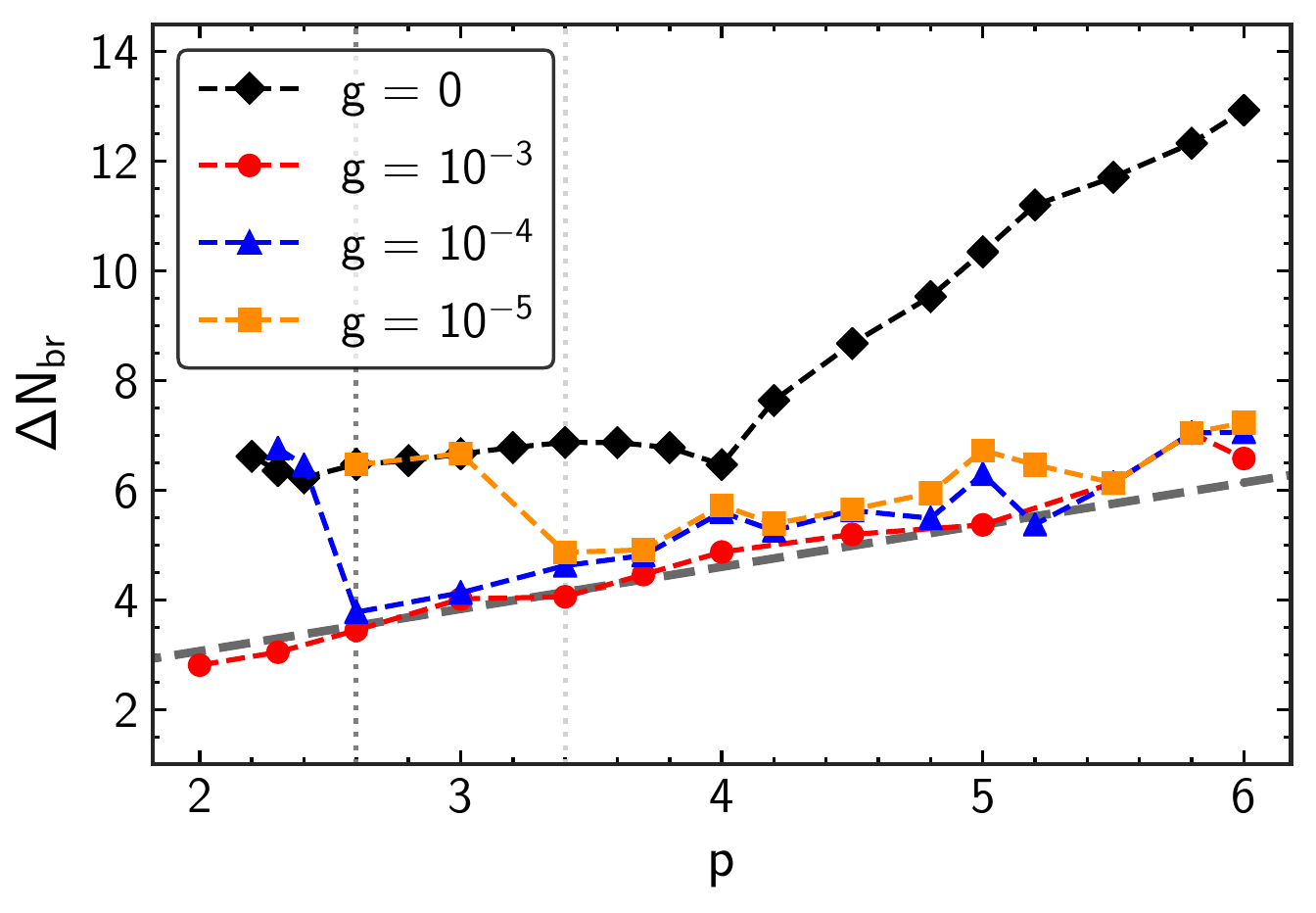} \,\,  \vspace*{-0.3cm}
    \caption{Number of e-folds $\Delta N_{\rm br}$ after inflation until $w$ deviates from $w_{\rm hom}$, due to backreaction.  The dashed gray line is the estimation $\Delta N_{\rm br} \simeq (p+2)\log(z_{\rm br})/6$ for $z_{\rm br} = 10^2$.} \vspace*{-0.3cm}
    \label{fig:PvsN-letter}
\end{figure}

\textbf{Inflationary observables and discussion -.} To compute the inflationary scalar tilt $n_s$ and tensor-to-scalar ratio $r$, we need to determine the number of e-folds $N_{\rm CMB}$ before the end of inflation, when the pivot scale $k_{\rm CMB} = 0.05 {\rm Mpc}^{-1}$ exited the horizon. For this, we need to know the exact evolution of the universe after inflation. In particular, we need \cite{Dodelson:2003vq,Liddle:2003as} \vspace*{-0.1cm}
\be
N_{\rm CMB}\simeq 61.5-\Delta N_\mathrm{br}+\mathrm{ln}\frac{V_{\rm CMB}^{1/2}}{m_\mathrm{pl}\rho_\mathrm{br}^{1/4}}+\frac{1-3\bar{w}}{12(1+\bar{w})}\mathrm{ln}\frac{\rho_\mathrm{RD}}{\rho_\mathrm{br}}, \label{eq:Nstar}
\ee
with $V_{\rm CMB}$ denoting the potential energy when $k_{\rm CMB}$ leaves the horizon, $\rho_\mathrm{br}$ and $\rho_\mathrm{RD}$ the energy densities when backreaction becomes noticeable and when the universe becomes RD, respectively, $\Delta N_{\rm br}$ the e-folds between the end of inflation and backreaction (see Fig.~\ref{fig:PvsN-letter}), and $\bar{w}$ the mean EoS  between backreaction and the onset of RD. For $p \gtrsim 3$, the transition from backreaction to RD is actually almost instantaneous, independently of $g^2$ (see Fig.~\ref{fig:EnergyRatios}), making the last term in Eq.~(\ref{eq:Nstar}) negligible. In Fig.~\ref{fig:PvsN-letter} we show $\Delta N_{\rm br}$ for different choices of $p$ and $g$, extracted from simulations. The inflaton fragments due to self-resonance for $p>2$ even if $g=0$, which provides the value for $\Delta N_{\rm br}$ found in~\cite{Lozanov:2016hid,Lozanov:2017hjm}. However, the presence of an interaction reduces significantly this quantity, as long as backreaction from the daughter field fragments the inflaton condensate. This requires e.g.~$p\gtrsim 3.4, 2.6$ for $g = 10^{-5},10^{-4}$ respectively.

For $p=2$, the system never achieves a RD state in our set-up, so we cannot determine $N_{\rm CMB}$. However, according to our results, the difference in $N_{\rm CMB}$ compared to a case in which the inflaton remains homogeneous until RD is $\delta N_{\rm CMB} \lesssim 1$ for $q_*<10^6$. For example, we get $\delta N_{\rm CMB} \approx 0.4$ for the case depicted in the top-left panel of Fig.~\ref{fig:EnergyRatios}. For $p>2$ we can compute $N_{\rm CMB}$ exactly, provided we note that $V_{\rm CMB}$ depends also on $N_{\rm CMB}$, making (\ref{eq:Nstar}) a non-linear equation.  
For $p=3-6$, we find the narrow range $N_{\rm CMB} \simeq 56.1-56.9$ for $M=4m_{\rm pl}$ and $N_{\rm CMB} \simeq 56.7-57.6$ for $M=10m_{\rm pl}$. Using this, we obtain very precise values for the inflationary observables:  $n_s \simeq 0.9643-0.9647$ and $r\simeq0.01-0.009$ for $M=4m_{\rm pl}$, and $n_s \simeq0.9633-0.9622$ and $r\simeq0.047-0.05$ for $M=10m_{\rm pl}$, see Fig.~\ref{fig:RNS-plane}. This represents a drastic reduction in the uncertainty of these quantities, compared to the traditional bounds obtained from $N_{\rm CMB} = 50-60$.

\textit{Discussion}. We have characterized in detail the evolution of the energy distribution and effective EoS of the universe from the end of inflation till the onset of RD, considering an inflaton with monomial potential during the (p)reheating stage, and a quadratic coupling to a light daughter field. Remarkable facts emerge:

\begin{figure}
    \begin{center}
    \includegraphics[height=5.1cm]{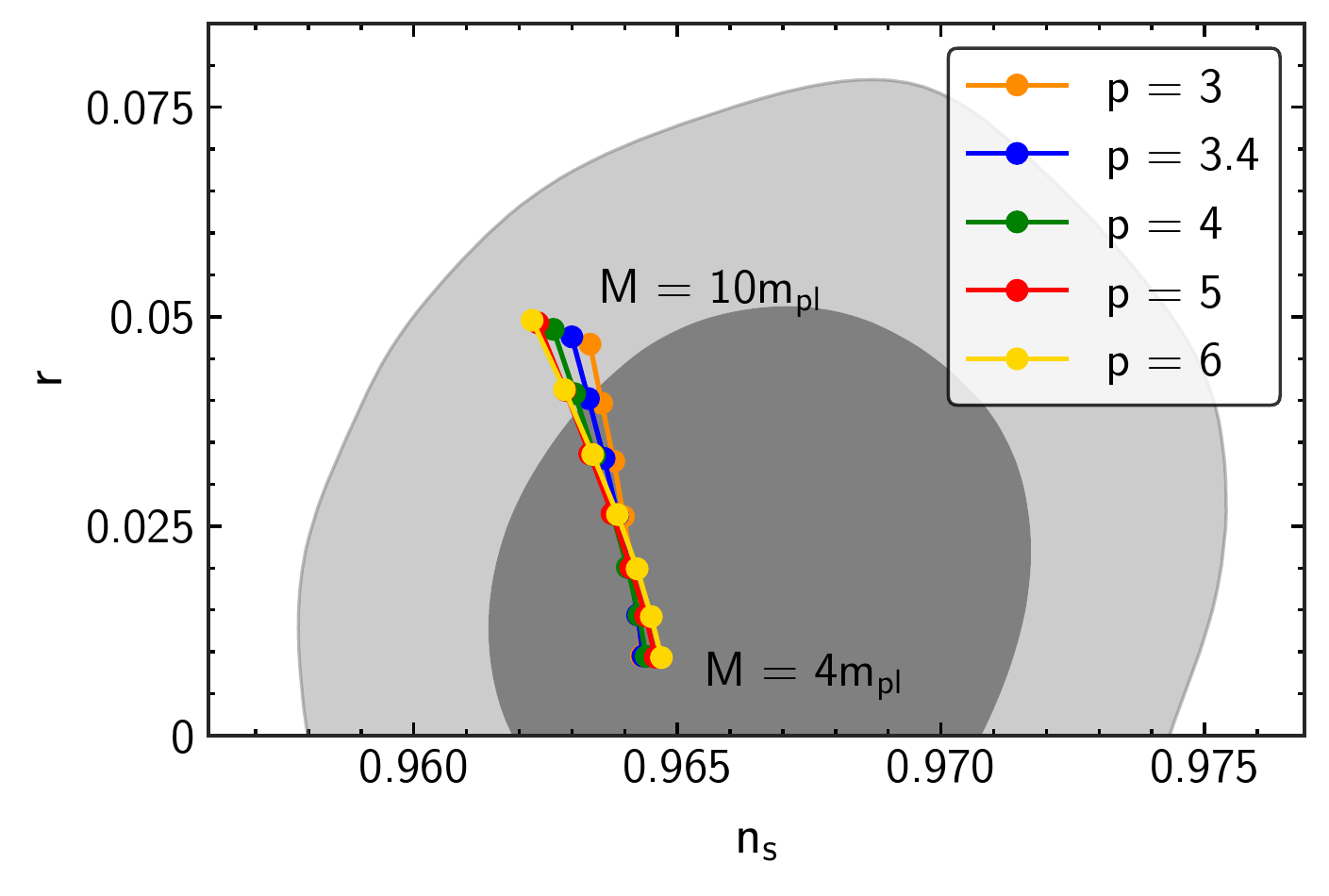} \vspace*{-0.3cm}
\end{center}
    \caption{Values of $n_s$ and $r$ predicted for different choices of $p$ and $M/m_{\rm pl} = 4-10$, indicated by dots. We take values satisfying $q_{*} > 10^{1.9 (4-p)}$ for $p<4$, and $q_* > 1$ for $p\geq4$. Contours show the observational constraints from Planck \cite{Akrami:2018odb}.
    } \vspace*{-0.5cm} \label{fig:RNS-plane}
\end{figure}

$i)$ Broad parametric resonance dominates over inflaton self-resonance, and backreaction from the daughter field is responsible for breaking the initial homogeneity of the inflaton. However, broad resonance eventually ends for $2 \leq p < 4$. For $p = 2$ the system goes back to a higher degree of homogeneity, while the EoS approaches gradually the homogeneous value ($w = 0$). For $p > 2$, inflaton fluctuations are also created via self-resonance, remarkably \textit{even after the breaking of the inflaton homogeneous condensate}. Due to this, the system always goes eventually to RD, either in the presence or absence of interactions with a daughter field species. 

$ii)$ The final amount of energy transferred to the daughter field is \textit{essentially independent of the coupling strength between the two fields}, and depends only on the power law exponent $p$: it becomes (eventually) negligible for $2 \leq p < 4$, and of order $\sim 50\%$ for $p \geq 4$. Therefore, in order to achieve a complete decay of the inflaton in these scenarios, some new ingredient is needed. 

$iii)$ Viable models of inflation with $p>2$ allow for a precise calculation of $N_{\rm CMB}$, with accuracy $\delta N_{\rm CMB} \lesssim 1$, leading to very precise predictions for $n_s$ and $r$. This highlights the relevance of characterizing the post-inflationary stage in detail. 

To conclude, we mention some limitations of our analysis, which can provide interesting avenues for future studies. For example, our study could be generalized to e.g.~trilinear interactions or higher order operators~\cite{Dufaux:2006ee}, or to an initial excitation via tachyonic preheating \cite{Felder:2000hj,Felder:2001kt,GarciaBellido:2002aj,Copeland:2002ku}. Oscillons can also form whenever the inflaton oscillates around flatter-than-quadratic regions of the potential, via self-resonance effects~\cite{Amin:2011hj} or tachyonic oscillations~\cite{Antusch:2015nla}. This would push the EoS towards $w \simeq 0$ \cite{Gleiser:2014ipa,Lozanov:2016hid,Lozanov:2017hjm} during their lifetime. 
While there are various effects that can change the early stages of preheating, we expect that the late-time energy distribution and EoS will depend mainly on the inflaton potential around its minimum, and on the type of inflaton-daughter coupling. Our analysis could also be generalized to multi-field inflation scenarios \cite{DeCross:2015uza,DeCross:2016fdz,DeCross:2016cbs,Krajewski:2018moi,Iarygina:2018kee}. In particular, the post-inflationary dynamics of a two-field inflation model with quartic potentials, non-minimal couplings, and quadratic interaction, has been studied with lattice simulations in \cite{Nguyen:2019kbm,vandeVis:2020qcp}, finding that RD is achieved in less than three e-folds for a significant fraction of the parameter space, in qualitative agreement with our results. Also, metric perturbations could be included \cite{Martin:2020fgl}.

Finally, if the inflaton is coupled to several light scalar fields, preliminary lattice simulations (for quadratic couplings) show that the energy transferred to the preheat sector can be enhanced up to ${N_f /(N_f + 1)}\%$, with $N_f$ being the number of different light daughter fields. We plan to explore some of these topics in the future.

\appendix

\section{} \label{app:1}

In this appendix we provide expressions for the field equations and energy components, and compare the outcome from (2+1)-dimensional and (3+1)-dimensional lattice simulations.
The equations of motion of the two fields ($f=\phi, X$) in a FLRW background, and of the background itself, are
\begin{eqnarray} \ddot{f} - a^{-2}\nabla_{\vec{x}}^2 f + 3H\dot{f}+\partial_f V=0 \,, \\
{\ddot a\over a} = {1\over 3m_{\rm pl}^2}\left\langle V(\phi,X) - {\dot \phi}^2 - {\dot X}^2\right\rangle \end{eqnarray}
where $V(\phi,X)$ is the potential given in Eq.~(\ref{eq:inflaton-potential}), $H \equiv \dot{a} /a$ is the Hubble rate, $a$ the scale factor, and $\langle ... \rangle$ stands for volume averaging. We define a new set of dimensionless field amplitudes and space-time variables by
\bea\label{eq:newvars1}
\varphi \equiv a^{\frac{6}{p+2}} (\phi / \phi_*) \ , \hspace{0.4cm} \chi \equiv a^{\frac{6}{p+2}} ( X  / \phi_* ) \ , \hspace{0.5cm}  \\
t \rightarrow z \equiv \int_{t_*}^t {\omega_* \, a(t')^{\frac{3 (2-p)}{p+2}}dt'} \ , \hspace{0.4cm} \vec{x} \rightarrow \vec{y} \equiv \omega_* \vec{x} \,, \hspace{0.5cm}
\eea
so that the period and amplitude of the inflaton oscillations are approximately constant and of order unity, and where
$\phi_*$ 
and $\omega_*$ are the initial inflaton amplitude and frequency at the end of inflation.
 In these variables, the equations of motion are
\bea\label{eq:fullEOMs1-l}
\hspace*{-0.4cm}\varphi'' - a^{\frac{-(16 - 4p)}{2+p}} \nabla^2_{\vec{y}} \varphi +  ( |\varphi|^{p-2} + q_{\rm res} \chi^2  + \Delta ) \varphi = 0  \ , \\
\hspace*{-0.4cm} \chi''- a^{\frac{-(16 - 4p)}{2+p}} \nabla^2_{\vec{y}} \chi + ( q_{\rm res} \varphi ^2 + \Delta ) \chi = 0 \ ,  \label{eq:fullEOMs2-l}   \eea
with $' \equiv  d/ dz$ and $\nabla_{\vec y} \equiv d / d \vec{y} $, and $q_{\rm res}$ the resonance parameter given in Eq.~(3) of the main text. Here, $\Delta \equiv \Delta (a'/a, a''/a)$  is the following time-dependent function \be \Delta \equiv \frac{6 (p-4)}{(p+2)^2} \left( \frac{a'}{a} \right)^2 + \frac{6}{p+2} \left( \frac{a''}{a} \right) \ . \ee
As the scale factor grows as $a \sim z^{\frac{p+2}{6}}$ during the initial linear regime of inflaton oscillations, $\Delta$ scales as $\sim z^{-2}$, and hence it becomes soon negligible, so it can be discarded for the following analysis. We can expand the fields up to linear order as $\varphi (\vec{y}, z) \equiv \bar{\varphi} (z) + \delta \varphi (\vec{y},z)$ and $\chi (\vec{y}, z) \equiv  \delta \chi (\vec{y},z)$, with the bar notation indicating the homogeneous/zero mode (the zero mode of the daughter field at the end of inflation is $\bar{\chi} (z)\simeq0$). From Eq.~(\ref{eq:fullEOMs1-l}) we get that the $eom$ of the inflaton zero mode is $\bar{\varphi}'' + |\bar{\varphi}|^{p-2}  \bar{\varphi} \simeq  0$, which gives rise to an oscillatory solution. On the other hand, the first order linearized equations for the modes $\delta\varphi_k$ and $\delta\chi_k$ are
\bea
 \delta\varphi_k'' +  ( \kappa^2+ (p-1)|\bar{\varphi }|^{p-2} ) \delta\varphi_k \simeq 0\,, \\ 
 \delta\chi_k''+ ( \kappa^2+  q_{\rm res} \bar{\varphi}^2  ) \delta\chi_k \simeq 0 \ , 
\eea
where $\kappa=a^{2(p - 4)\over p+2}k/\omega_*$. These modes have time-dependent effective frequencies,
leading to solutions as $|\delta \chi_k |^2 \propto e^{2\mu_k z}$ and $|\delta \varphi_k |^2 \propto e^{2\nu_k z}$, with $\mu_k \equiv \mu_k (\kappa,q_{\rm res};p)$ and $\nu_k \equiv \nu_k (\kappa;p)$ their respective {\it Floquet} indices. The parametric resonance regime is characterized by the exponentially growing solutions obtained when the Floquet index becomes a positive number within a range of momenta. 

The energy and pressure densities of the fields are
\begin{alignat*}{3}
 \rho &\,\,=\,\,&& \frac{1}{2} \dot{\phi}^2 +  \frac{1}{2} \dot{X}^2  + \frac{1}{2} |\nabla \phi |^2 + \frac{1}{2} |\nabla X |^2 + V (\phi, X) \\ 
&\,\,=\,\,&& \frac{\omega_*^2 \phi_*^2}{a^{\frac{6p}{2+p}} } ( E_{\rm k, \varphi} + E_{\rm k, \chi} + E_{\rm g, \varphi}  + E_{\rm g, \chi}+ E_{\rm int} + E_{\rm pot} ) \,, \\ 
p &\,\,=\,\,&& \frac{1}{2} \dot{\phi}^2 +  \frac{1}{2} \dot{X}^2 - \frac{1}{6} |\nabla \phi |^2 - \frac{1}{6} |\nabla X |^2 - V (\phi, X)  \\ 
&\,\,=\,\,&& \frac{\omega_*^2 \phi_*^2}{a^{\frac{6p}{2+p}} }  ( E_{\rm k, \varphi} + E_{\rm k, \chi} - \frac{1}{3} E_{\rm g, \varphi}  - \frac{1}{3} E_{\rm g, \chi} - E_{\rm int} - E_{\rm pot} ) \,, 
\end{alignat*}
where the subindex `k' refers to the kinetic energy of any of the two fields, `g' refers to their gradient energy,  `int' to the interaction energy, and `pot' to the inflaton potential energy. The form of these terms is the following ($f = \varphi,\chi$),
\bea\label{eq:energy1} 
E_{\rm k, f} &\equiv &   \frac{1}{2} \left( f' - \frac{6}{p+2} \frac{a'}{a} f \right)^2 \ , \hspace{0.4cm} E_{\rm pot} \equiv   \frac{1}{p} \varphi^p  \,, \nonumber \\ 
E_{\rm int} &\equiv & \frac{1}{2} a^{\frac{6p-24}{p+2}} q_* \varphi^2 \chi^2 \ , \hspace{0.4cm} E_{\rm g, f} \equiv \frac{1}{2}  a^{\frac{4p-16}{p+2}} |\nabla_{\vec{y}} f |^2\nonumber \ . \eea

Lattice simulations have been carried out with the public package \texttt{Clustereasy} \cite{Felder:2007nz} (the MPI version of \texttt{Latticeeasy} \cite{Felder:2000hq}), which uses a second-order leapfrog-like algorithm. As a cross-check, some of the simulations have been repeated with velocity-verlet algorithms of higher order  implemented in \CLns, a recent package that was developed shortly after the completion of the bulk of the computations presented in this Letter \cite{Figueroa:2020rrl}. We have carried out lattice simulations of the field system in 2- and 3-spatial dimensions (2D and 3D from now on), obtaining that the dynamics are very similar at the quantitative level. Although 3D simulations are a better approximation to the field dynamics in the continuum, 2D simulations require significantly less computing time, allowing us to explore the very late-time regime of the system, as well as to increase the spatial resolution of the lattice whenever needed.

\begin{figure*}
    \centering
    \includegraphics[width=7.5cm]{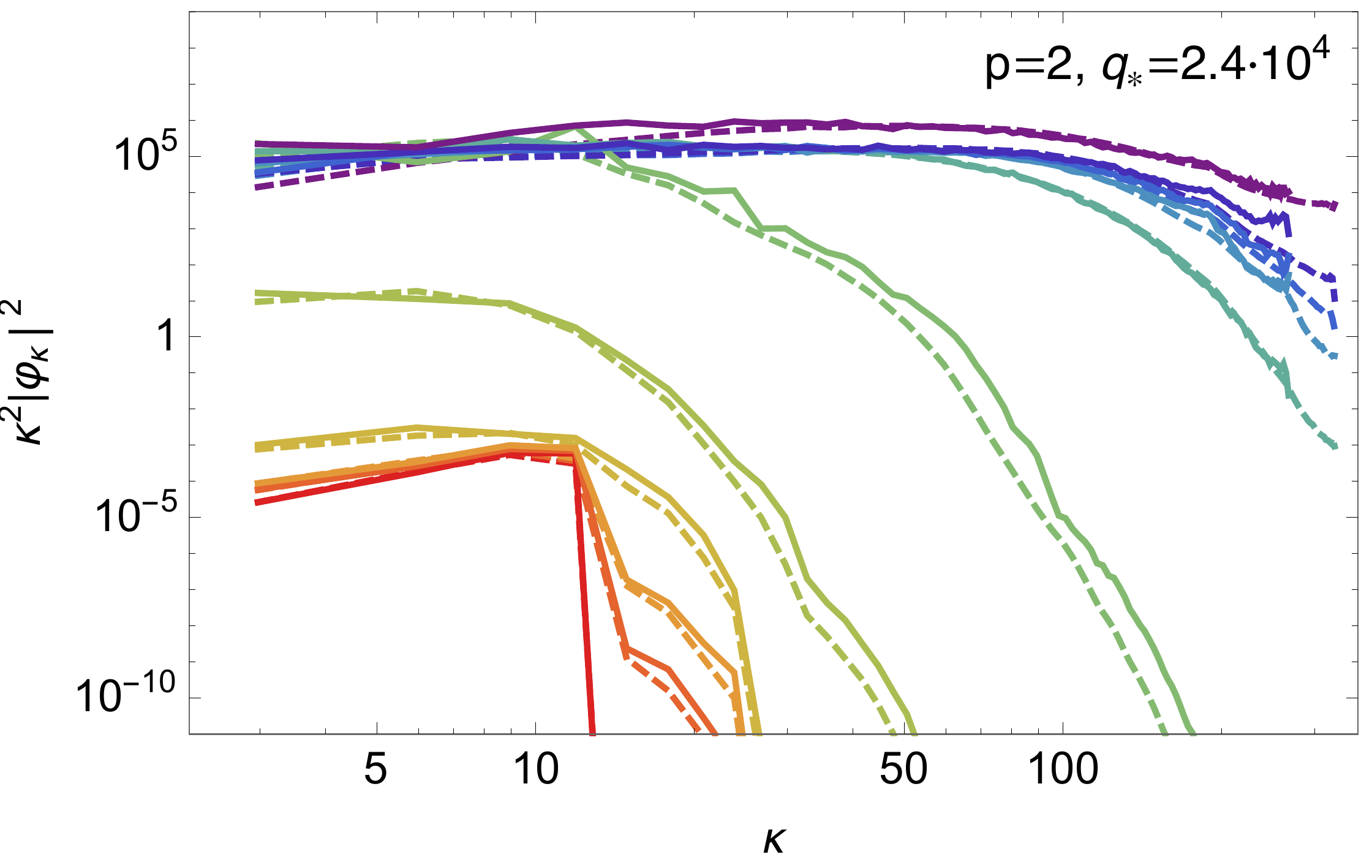} \hspace{0.5cm}
    \includegraphics[width=7.5cm]{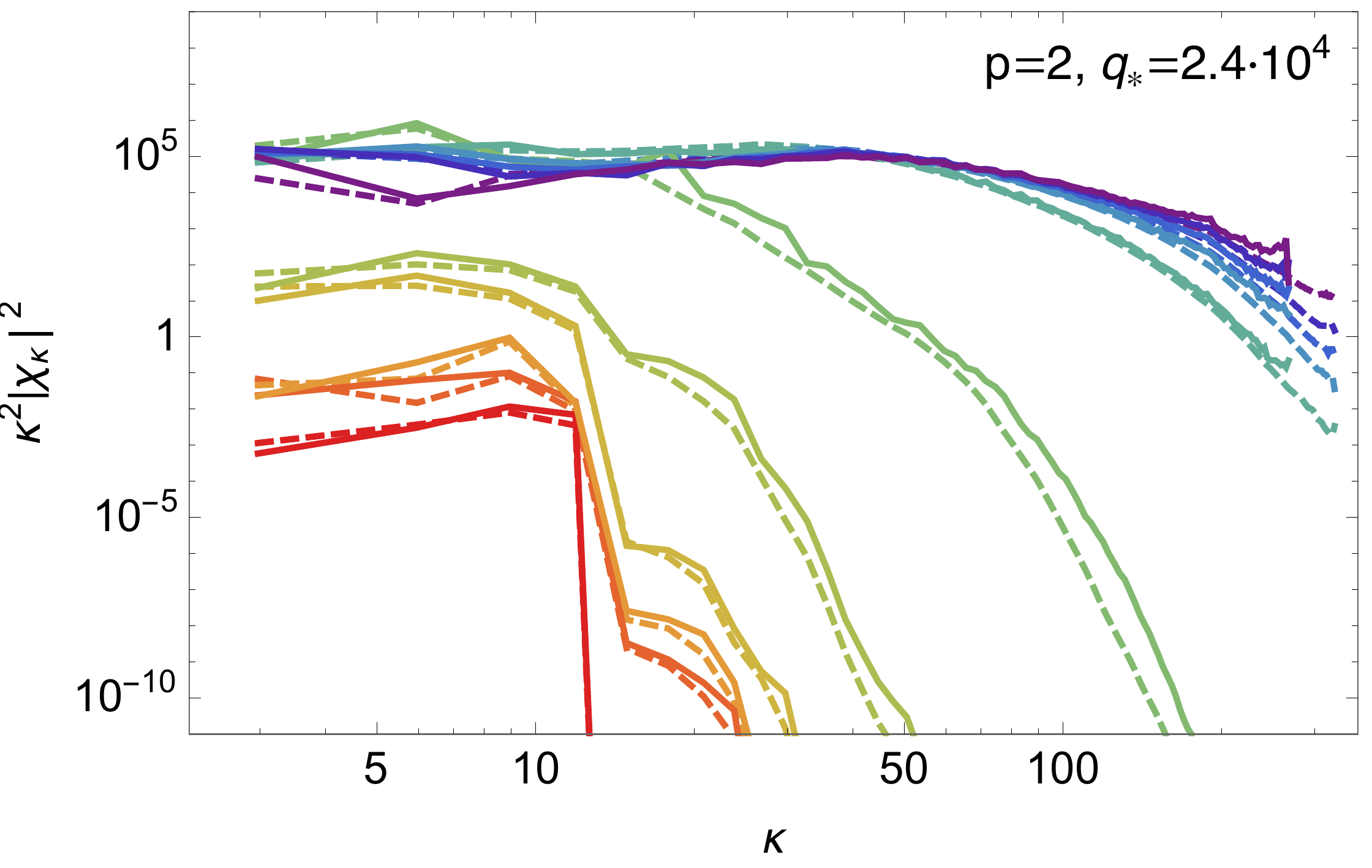} \vspace*{0.1cm} \\
    \includegraphics[width=7.5cm]{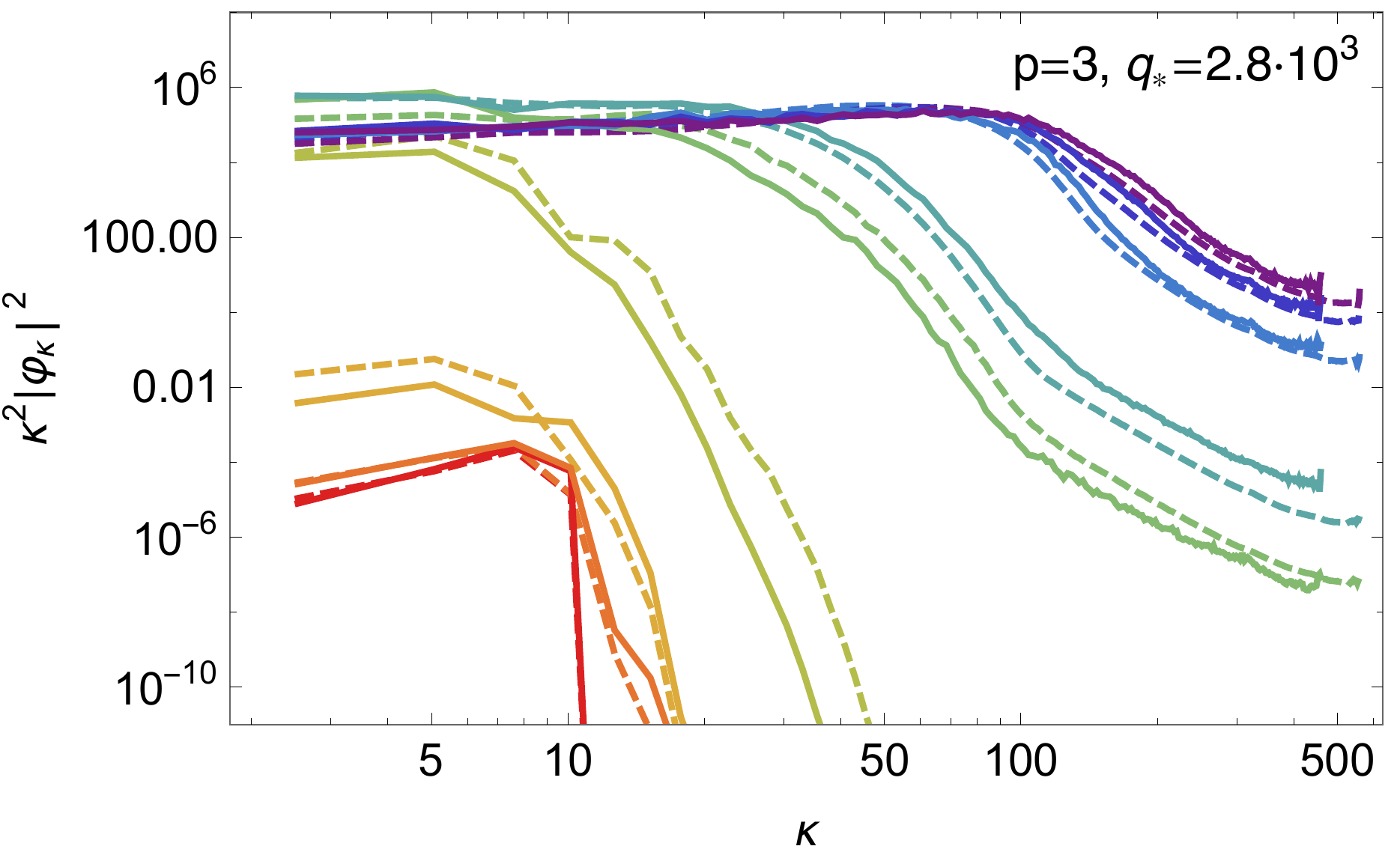} \hspace{0.5cm}
    \includegraphics[width=7.5cm]{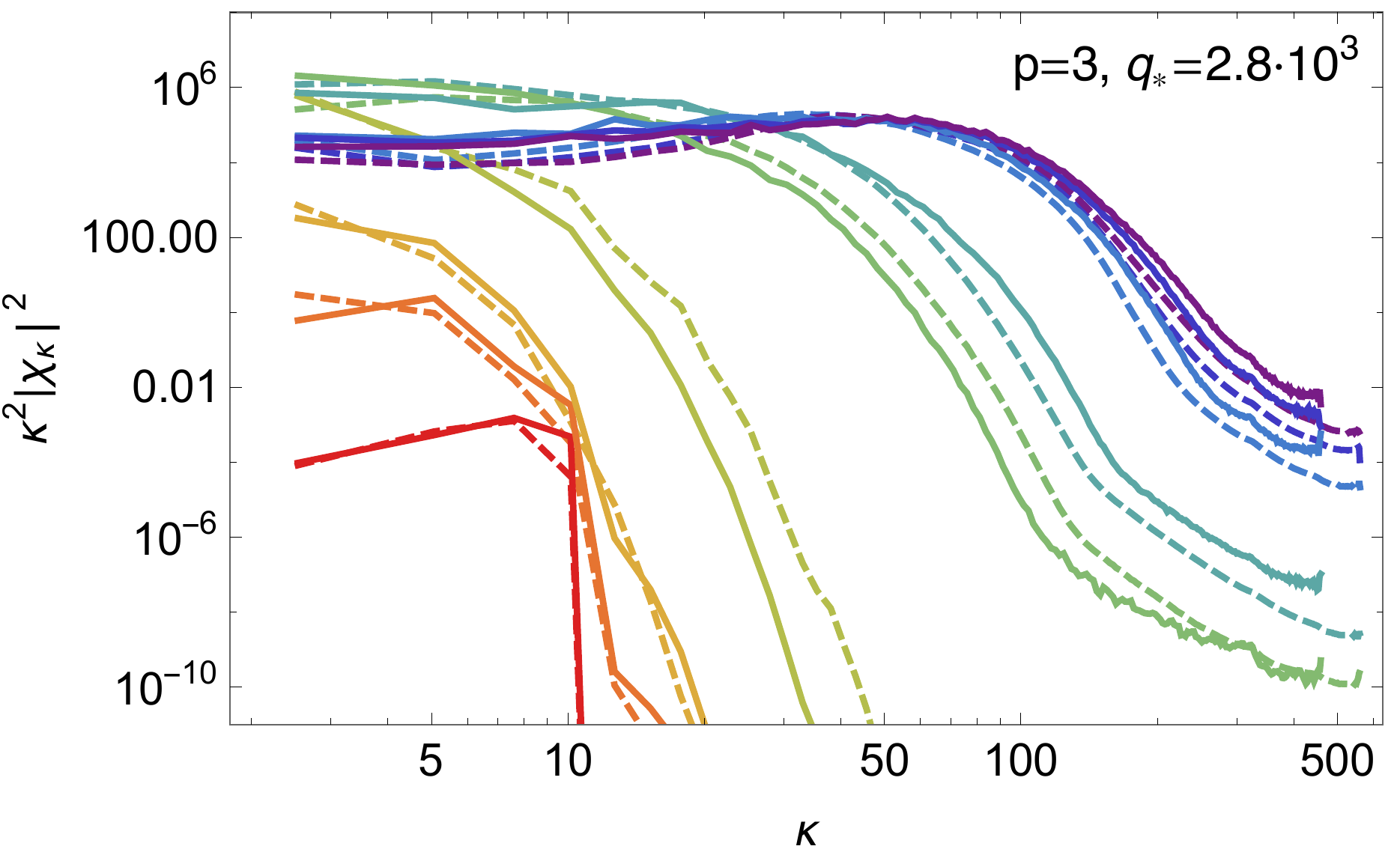}  \vspace*{0.1cm} \\
    \includegraphics[width=7.5cm]{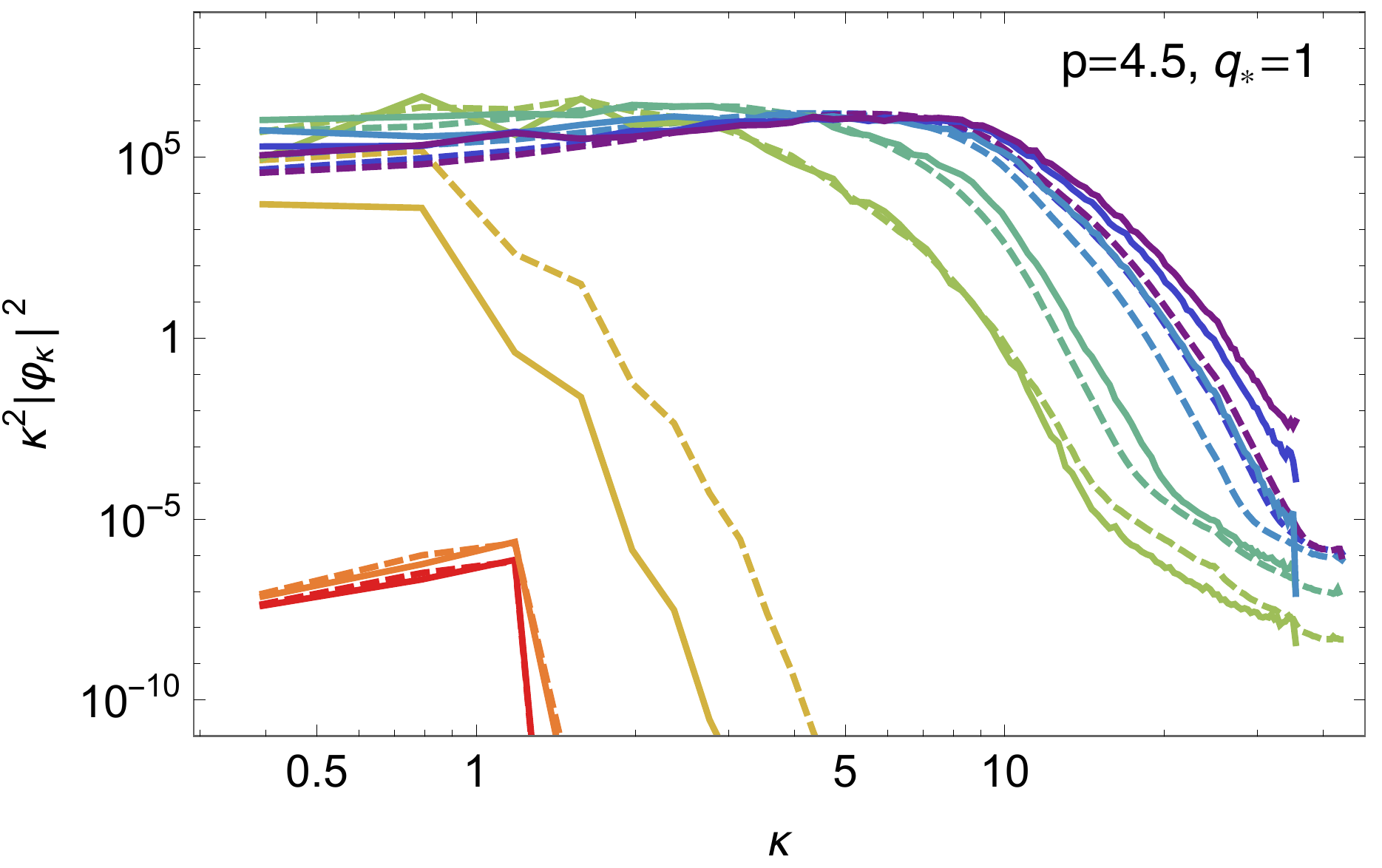} \hspace{0.5cm}
    \includegraphics[width=7.5cm]{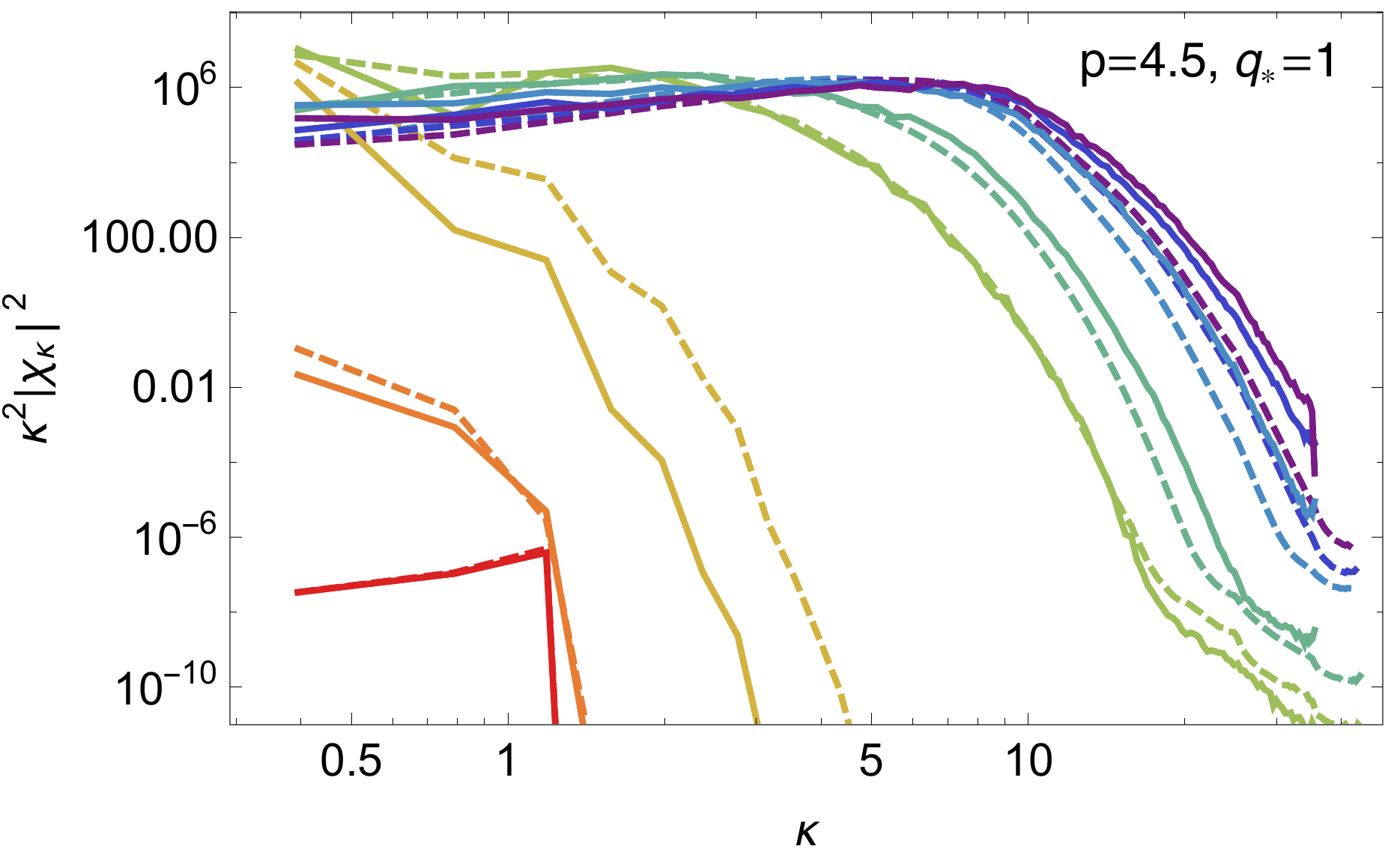} 
\caption{Comparison of the inflaton and daughter field spectra obtained with lattice simulations in 2+1 dimensions (continuous lines) and 3+1 dimensions (dashed lines), for three different choices of $p$ and $q_*$. Momenta are defined as $\kappa \equiv k / \omega_*$. Each colored line corresponds to a different time, going from red (early times) to purple (late times). } \label{fig:spectra-comparison}
\end{figure*}

\begin{figure*}
    \centering
    \includegraphics[width=5.3cm]{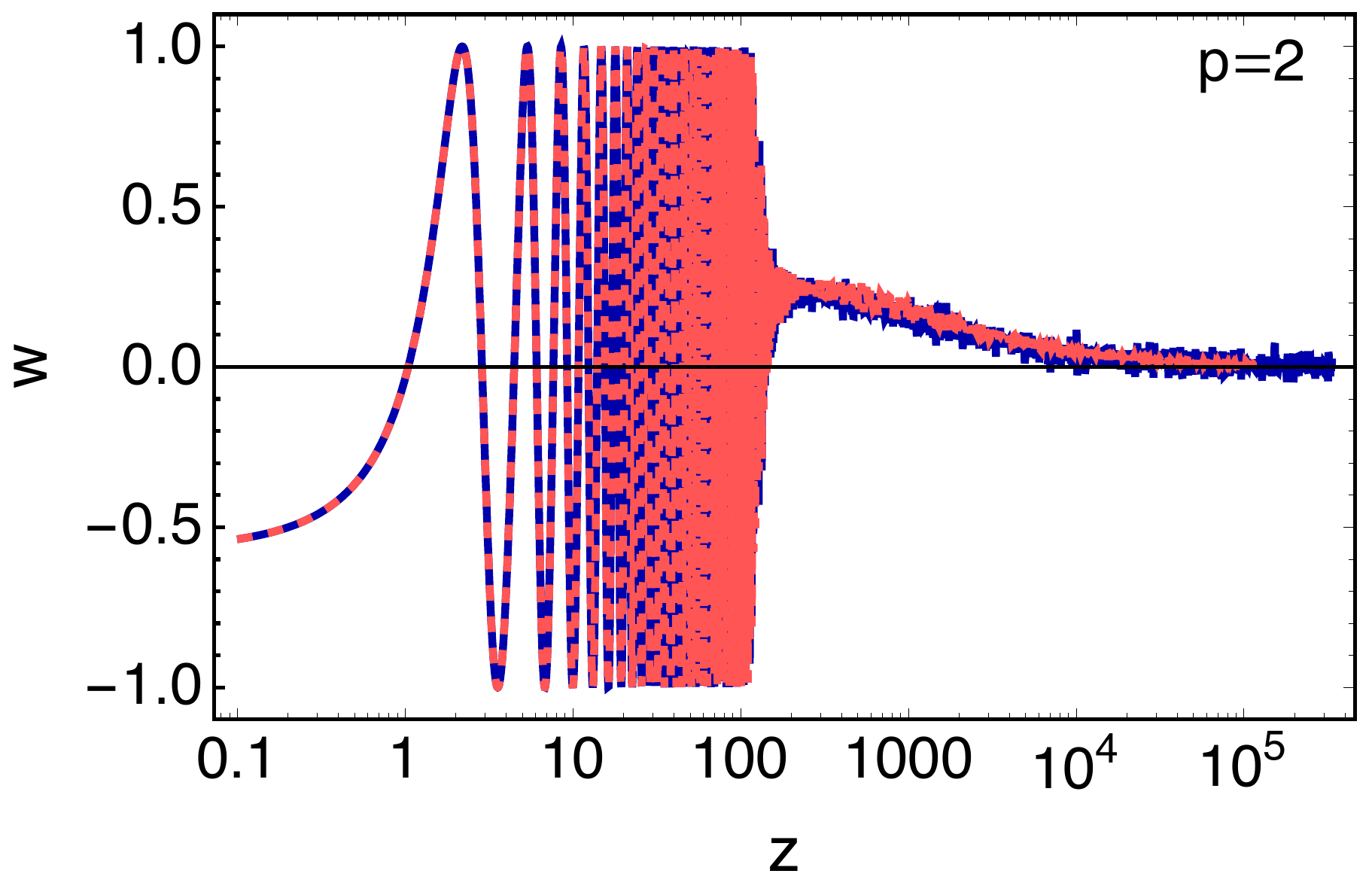} \hspace{0.1cm}
    \includegraphics[width=5.3cm]{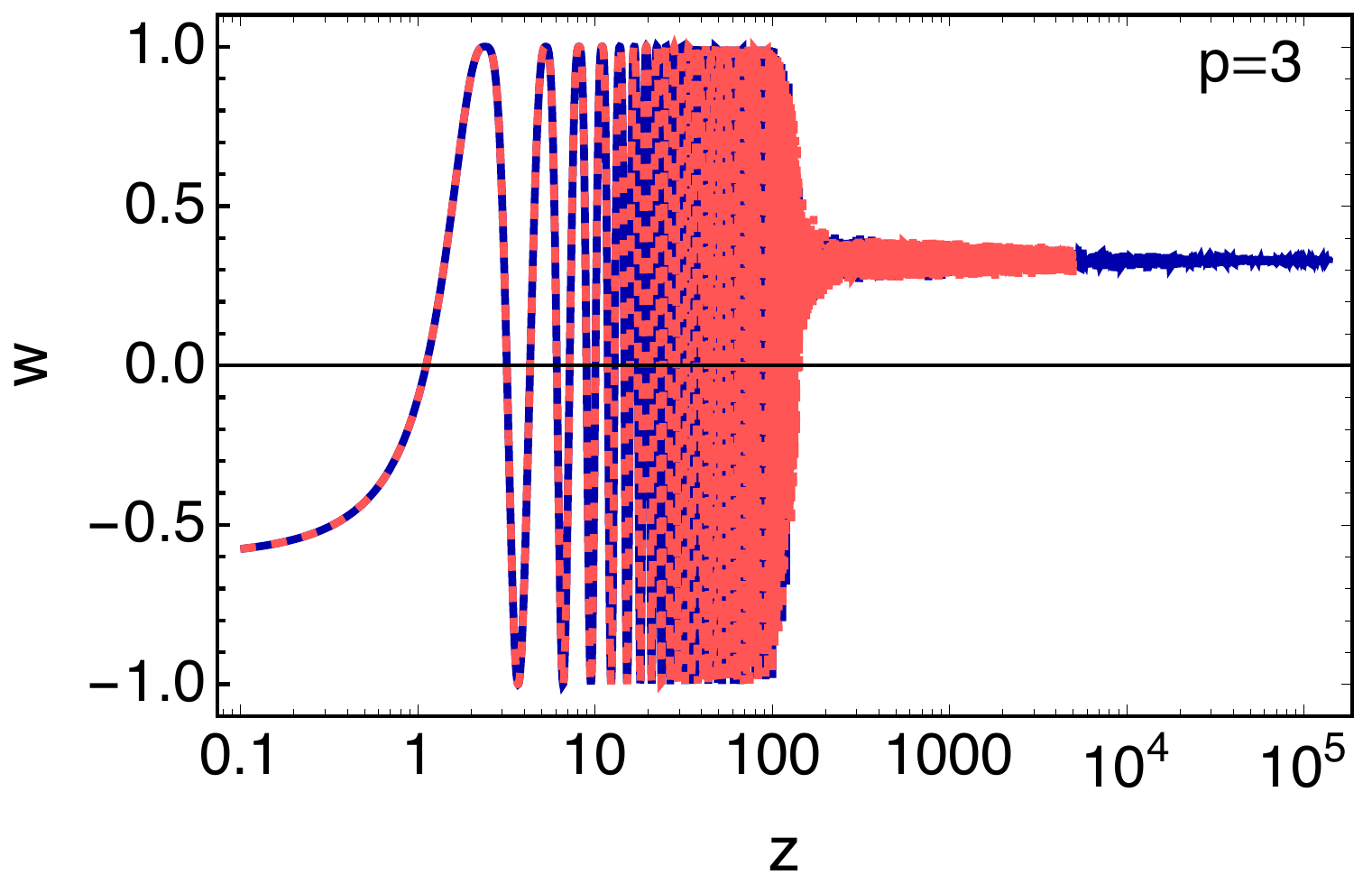} \hspace{0.1cm}
    \includegraphics[width=5.3cm]{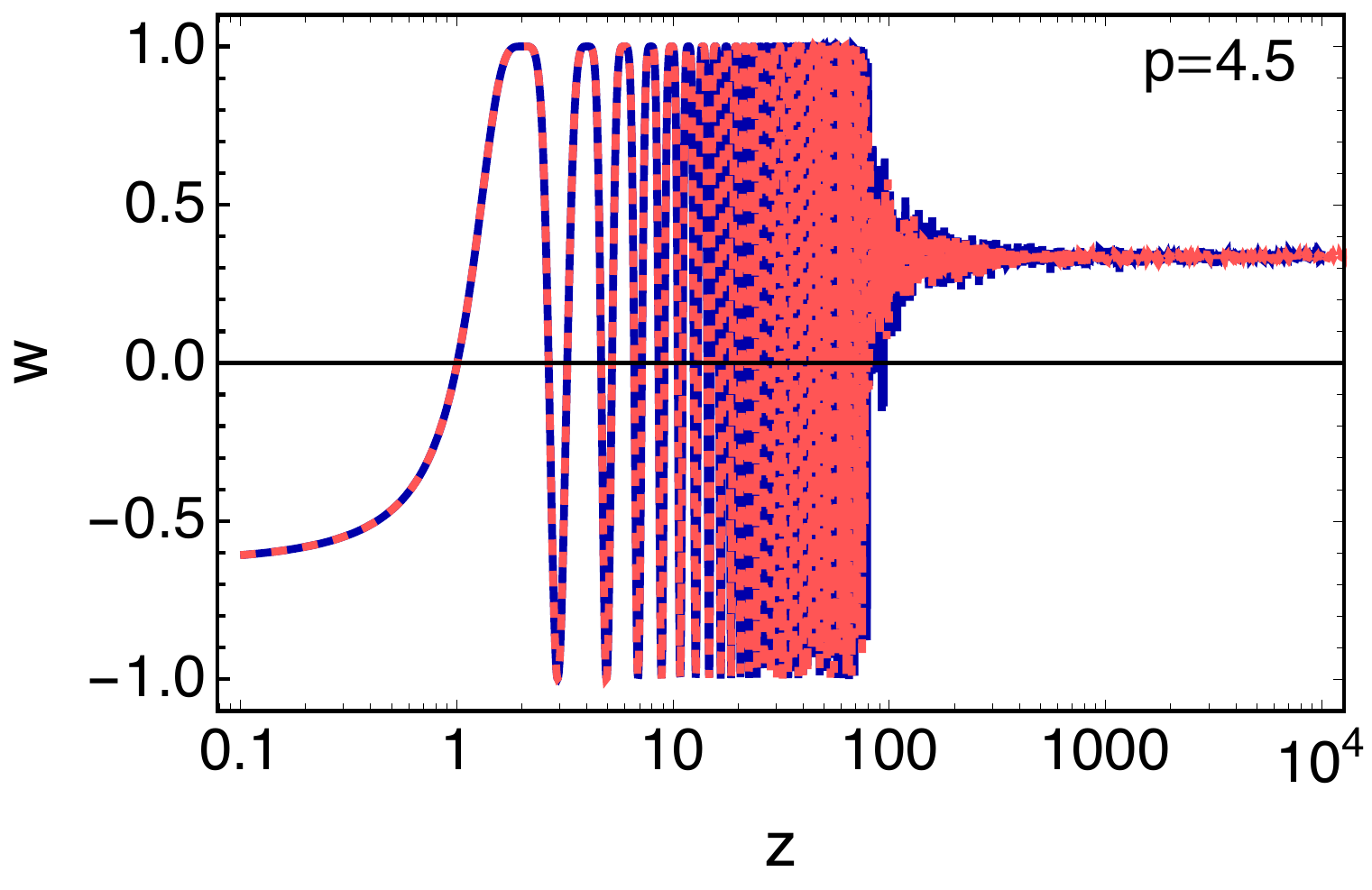}
    \caption{Comparison of the equation of state obtained with lattice simulations in 2+1 dimensions (blue line) and 3+1 dimensions (red dashed line), for $p=2$, 3, 4.5.  } \label{fig:eos-comparison}
\end{figure*}

We show in Fig.~\ref{fig:spectra-comparison} a direct comparison of the spectra of the inflaton and daughter field in 2D and 3D. For illustrative purposes, we have chosen the power-law coefficients $p= 2$, 3, and 4.5, which cover the three different dynamical regimes described in the paper. In order to do an appropriate comparison, the number of points per dimension and infrared cutoff of the lattice are the same in both simulations, which implies that the UV coverage in the 3D case is a factor $\sqrt{3/2}$ larger than in 2D. It can be clearly appreciated that the spectral evolution is very similar in the three depicted cases, both during the initial linear excitation regime, as well as during the later non-linear regime. We also show in Fig.~\ref{fig:eos-comparison} a direct comparison of the equation of state as a function of time, for the same set of model parameters. Its behaviour is identical in 2D and 3D simulations, in the three cases depicted here: backreaction time happens at approximately the same time, and the final value of the equation of state at asymptotically late times is similar: it goes to $w \rightarrow 0$ for $p=2$, and to $w \rightarrow 1/3$ for $p>2$. We have checked this for many other model parameters beyond the examples shown here. Our comparison results show that the use of 2D lattice simulations to parametrize the dynamics of the system is completely justified in our case of study.

\section*{Acknowledgements} DGF (ORCID 0000-0002-4005-8915) is supported by a Ram\'on y Cajal contract by Spanish Ministry MINECO, with Ref. RYC-2017-23493, and by the grant "SOM: Sabor y Origen de la Materia”, from Spanish Ministry of Science and Innovation, under no. FPA2017-85985-P. DGF and FT, acknowledge hospitality and support from KITP-UCSB, where part of this work was carried out, supported in part by the National Science Foundation Grant No.\ NSF PHY-1748958. S.~Antusch, K.~Marschall and F.~Torrenti acknowledge support from the Swiss National Science Foundation (project number 200020/175502).

\bibliography{EOSletter.bib}

\begin{thebibliography}{10}
\expandafter\ifx\csname url\endcsname\relax
  \def\url#1{\texttt{#1}}\fi
\expandafter\ifx\csname urlprefix\endcsname\relax\def\urlprefix{URL }\fi
\expandafter\ifx\csname href\endcsname\relax
  \def\href#1#2{#2} \def\path#1{#1}\fi

\bibitem{Starobinsky:1980te}
A.~A. Starobinsky, {A New Type of Isotropic Cosmological Models Without
  Singularity}, Phys. Lett. 91B (1980) 99--102, [Adv. Ser. Astrophys.
  Cosmol.3,130(1987)].
\newblock \href {https://doi.org/10.1016/0370-2693(80)90670-X}
  {\path{doi:10.1016/0370-2693(80)90670-X}}.

\bibitem{Guth:1980zm}
A.~H. Guth, {The Inflationary Universe: A Possible Solution to the Horizon and
  Flatness Problems}, Phys. Rev. D23 (1981) 347--356, [Adv. Ser. Astrophys.
  Cosmol.3,139(1987)].
\newblock \href {https://doi.org/10.1103/PhysRevD.23.347}
  {\path{doi:10.1103/PhysRevD.23.347}}.

\bibitem{Linde:1981mu}
A.~D. Linde, {A New Inflationary Universe Scenario: A Possible Solution of the
  Horizon, Flatness, Homogeneity, Isotropy and Primordial Monopole Problems},
  Phys. Lett. 108B (1982) 389--393, [Adv. Ser. Astrophys. Cosmol.3,149(1987)].
\newblock \href {https://doi.org/10.1016/0370-2693(82)91219-9}
  {\path{doi:10.1016/0370-2693(82)91219-9}}.

\bibitem{Albrecht:1982wi}
A.~Albrecht, P.~J. Steinhardt, {Cosmology for Grand Unified Theories with
  Radiatively Induced Symmetry Breaking}, Phys. Rev. Lett. 48 (1982)
  1220--1223, [Adv. Ser. Astrophys. Cosmol.3,158(1987)].
\newblock \href {https://doi.org/10.1103/PhysRevLett.48.1220}
  {\path{doi:10.1103/PhysRevLett.48.1220}}.

\bibitem{Akrami:2018odb}
Y.~Akrami, et~al., {Planck 2018 results. X. Constraints on inflation} (2018).
\newblock \href {http://arxiv.org/abs/1807.06211} {\path{arXiv:1807.06211}}.

\bibitem{Ade:2018gkx}
P.~A.~R. Ade, et~al., {BICEP2 / Keck Array x: Constraints on Primordial
  Gravitational Waves using Planck, WMAP, and New BICEP2/Keck Observations
  through the 2015 Season}, Phys. Rev. Lett. 121 (2018) 221301.
\newblock \href {http://arxiv.org/abs/1810.05216} {\path{arXiv:1810.05216}},
  \href {https://doi.org/10.1103/PhysRevLett.121.221301}
  {\path{doi:10.1103/PhysRevLett.121.221301}}.

\bibitem{Bassett:2005xm}
B.~A. Bassett, S.~Tsujikawa, D.~Wands, {Inflation dynamics and reheating}, Rev.
  Mod. Phys. 78 (2006) 537--589.
\newblock \href {http://arxiv.org/abs/astro-ph/0507632}
  {\path{arXiv:astro-ph/0507632}}, \href
  {https://doi.org/10.1103/RevModPhys.78.537}
  {\path{doi:10.1103/RevModPhys.78.537}}.

\bibitem{Allahverdi:2010xz}
R.~Allahverdi, R.~Brandenberger, F.-Y. Cyr-Racine, A.~Mazumdar, {Reheating in
  Inflationary Cosmology: Theory and Applications}, Ann. Rev. Nucl. Part. Sci.
  60 (2010) 27--51.
\newblock \href {http://arxiv.org/abs/1001.2600} {\path{arXiv:1001.2600}},
  \href {https://doi.org/10.1146/annurev.nucl.012809.104511}
  {\path{doi:10.1146/annurev.nucl.012809.104511}}.

\bibitem{Amin:2014eta}
M.~A. Amin, M.~P. Hertzberg, D.~I. Kaiser, J.~Karouby, {Nonperturbative
  Dynamics Of Reheating After Inflation: A Review}, Int. J. Mod. Phys. D24
  (2014) 1530003.
\newblock \href {http://arxiv.org/abs/1410.3808} {\path{arXiv:1410.3808}},
  \href {https://doi.org/10.1142/S0218271815300037}
  {\path{doi:10.1142/S0218271815300037}}.

\bibitem{Lozanov:2019jxc}
K.~D. Lozanov, {Lectures on Reheating after Inflation} (2019).
\newblock \href {http://arxiv.org/abs/1907.04402} {\path{arXiv:1907.04402}}.

\bibitem{Allahverdi:2020bys}
R.~Allahverdi, et~al., {The First Three Seconds: a Review of Possible Expansion
  Histories of the Early Universe} (6 2020).
\newblock \href {http://arxiv.org/abs/2006.16182} {\path{arXiv:2006.16182}}.

\bibitem{Dai:2014jja}
L.~Dai, M.~Kamionkowski, J.~Wang, {Reheating constraints to inflationary
  models}, Phys. Rev. Lett. 113 (2014) 041302.
\newblock \href {http://arxiv.org/abs/1404.6704} {\path{arXiv:1404.6704}},
  \href {https://doi.org/10.1103/PhysRevLett.113.041302}
  {\path{doi:10.1103/PhysRevLett.113.041302}}.

\bibitem{Martin:2014nya}
J.~Martin, C.~Ringeval, V.~Vennin, {Observing Inflationary Reheating}, Phys.
  Rev. Lett. 114~(8) (2015) 081303.
\newblock \href {http://arxiv.org/abs/1410.7958} {\path{arXiv:1410.7958}},
  \href {https://doi.org/10.1103/PhysRevLett.114.081303}
  {\path{doi:10.1103/PhysRevLett.114.081303}}.

\bibitem{Munoz:2014eqa}
J.~B. Munoz, M.~Kamionkowski, {Equation-of-State Parameter for Reheating},
  Phys. Rev. D91~(4) (2015) 043521.
\newblock \href {http://arxiv.org/abs/1412.0656} {\path{arXiv:1412.0656}},
  \href {https://doi.org/10.1103/PhysRevD.91.043521}
  {\path{doi:10.1103/PhysRevD.91.043521}}.

\bibitem{Gong:2015qha}
J.-O. Gong, S.~Pi, G.~Leung, {Probing reheating with primordial spectrum}, JCAP
  05 (2015) 027.
\newblock \href {http://arxiv.org/abs/1501.03604} {\path{arXiv:1501.03604}},
  \href {https://doi.org/10.1088/1475-7516/2015/05/027}
  {\path{doi:10.1088/1475-7516/2015/05/027}}.

\bibitem{Kallosh:2013hoa}
R.~Kallosh, A.~Linde, {Universality Class in Conformal Inflation}, JCAP 1307
  (2013) 002.
\newblock \href {http://arxiv.org/abs/1306.5220} {\path{arXiv:1306.5220}},
  \href {https://doi.org/10.1088/1475-7516/2013/07/002}
  {\path{doi:10.1088/1475-7516/2013/07/002}}.

\bibitem{Figueroa:2015rqa}
D.~G. Figueroa, J.~Garcia-Bellido, F.~Torrenti, {Decay of the standard model
  Higgs field after inflation}, Phys. Rev. D92~(8) (2015) 083511.
\newblock \href {http://arxiv.org/abs/1504.04600} {\path{arXiv:1504.04600}},
  \href {https://doi.org/10.1103/PhysRevD.92.083511}
  {\path{doi:10.1103/PhysRevD.92.083511}}.

\bibitem{Traschen:1990sw}
J.~H. Traschen, R.~H. Brandenberger, {Particle Production During
  Out-of-equilibrium Phase Transitions}, Phys. Rev. D42 (1990) 2491--2504.
\newblock \href {https://doi.org/10.1103/PhysRevD.42.2491}
  {\path{doi:10.1103/PhysRevD.42.2491}}.

\bibitem{Kofman:1994rk}
L.~Kofman, A.~D. Linde, A.~A. Starobinsky, {Reheating after inflation}, Phys.
  Rev. Lett. 73 (1994) 3195--3198.
\newblock \href {http://arxiv.org/abs/hep-th/9405187}
  {\path{arXiv:hep-th/9405187}}, \href
  {https://doi.org/10.1103/PhysRevLett.73.3195}
  {\path{doi:10.1103/PhysRevLett.73.3195}}.

\bibitem{Shtanov:1994ce}
Y.~Shtanov, J.~H. Traschen, R.~H. Brandenberger, {Universe reheating after
  inflation}, Phys. Rev. D51 (1995) 5438--5455.
\newblock \href {http://arxiv.org/abs/hep-ph/9407247}
  {\path{arXiv:hep-ph/9407247}}, \href
  {https://doi.org/10.1103/PhysRevD.51.5438}
  {\path{doi:10.1103/PhysRevD.51.5438}}.

\bibitem{Khlebnikov:1996mc}
S.~{\relax Yu}. Khlebnikov, I.~I. Tkachev, {Classical decay of inflaton}, Phys.
  Rev. Lett. 77 (1996) 219--222.
\newblock \href {http://arxiv.org/abs/hep-ph/9603378}
  {\path{arXiv:hep-ph/9603378}}, \href
  {https://doi.org/10.1103/PhysRevLett.77.219}
  {\path{doi:10.1103/PhysRevLett.77.219}}.

\bibitem{Prokopec:1996rr}
T.~Prokopec, T.~G. Roos, {Lattice study of classical inflaton decay}, Phys.
  Rev. D55 (1997) 3768--3775.
\newblock \href {http://arxiv.org/abs/hep-ph/9610400}
  {\path{arXiv:hep-ph/9610400}}, \href
  {https://doi.org/10.1103/PhysRevD.55.3768}
  {\path{doi:10.1103/PhysRevD.55.3768}}.

\bibitem{Kofman:1997yn}
L.~Kofman, A.~D. Linde, A.~A. Starobinsky, {Towards the theory of reheating
  after inflation}, Phys. Rev. D56 (1997) 3258--3295.
\newblock \href {http://arxiv.org/abs/hep-ph/9704452}
  {\path{arXiv:hep-ph/9704452}}, \href
  {https://doi.org/10.1103/PhysRevD.56.3258}
  {\path{doi:10.1103/PhysRevD.56.3258}}.

\bibitem{Greene:1997fu}
P.~B. Greene, L.~Kofman, A.~D. Linde, A.~A. Starobinsky, {Structure of
  resonance in preheating after inflation}, Phys. Rev. D56 (1997) 6175--6192.
\newblock \href {http://arxiv.org/abs/hep-ph/9705347}
  {\path{arXiv:hep-ph/9705347}}, \href
  {https://doi.org/10.1103/PhysRevD.56.6175}
  {\path{doi:10.1103/PhysRevD.56.6175}}.

\bibitem{Lozanov:2016hid}
K.~D. Lozanov, M.~A. Amin, {Equation of State and Duration to Radiation
  Domination after Inflation}, Phys. Rev. Lett. 119~(6) (2017) 061301.
\newblock \href {http://arxiv.org/abs/1608.01213} {\path{arXiv:1608.01213}},
  \href {https://doi.org/10.1103/PhysRevLett.119.061301}
  {\path{doi:10.1103/PhysRevLett.119.061301}}.

\bibitem{Figueroa:2016wxr}
D.~G. Figueroa, F.~Torrenti, {Parametric resonance in the early Universe—a
  fitting analysis}, JCAP 1702 (2017) 001.
\newblock \href {http://arxiv.org/abs/1609.05197} {\path{arXiv:1609.05197}},
  \href {https://doi.org/10.1088/1475-7516/2017/02/001}
  {\path{doi:10.1088/1475-7516/2017/02/001}}.

\bibitem{Lozanov:2017hjm}
K.~D. Lozanov, M.~A. Amin, {Self-resonance after inflation: oscillons,
  transients and radiation domination}, Phys. Rev. D97~(2) (2018) 023533.
\newblock \href {http://arxiv.org/abs/1710.06851} {\path{arXiv:1710.06851}},
  \href {https://doi.org/10.1103/PhysRevD.97.023533}
  {\path{doi:10.1103/PhysRevD.97.023533}}.

\bibitem{Giblin:2019nuv}
J.~T. Giblin, A.~J. Tishue, {Preheating in Full General Relativity}, Phys. Rev.
  D100~(6) (2019) 063543.
\newblock \href {http://arxiv.org/abs/1907.10601} {\path{arXiv:1907.10601}},
  \href {https://doi.org/10.1103/PhysRevD.100.063543}
  {\path{doi:10.1103/PhysRevD.100.063543}}.

\bibitem{Figueroa:2018twl}
D.~G. Figueroa, E.~H. Tanin, {Inconsistency of an inflationary sector coupled
  only to Einstein gravity}, JCAP 1910 (2019) 050.
\newblock \href {http://arxiv.org/abs/1811.04093} {\path{arXiv:1811.04093}},
  \href {https://doi.org/10.1088/1475-7516/2019/10/050}
  {\path{doi:10.1088/1475-7516/2019/10/050}}.

\bibitem{Figueroa:2019paj}
D.~G. Figueroa, E.~H. Tanin, {Ability of LIGO and LISA to probe the equation of
  state of the early Universe}, JCAP 1908 (2019) 011.
\newblock \href {http://arxiv.org/abs/1905.11960} {\path{arXiv:1905.11960}},
  \href {https://doi.org/10.1088/1475-7516/2019/08/011}
  {\path{doi:10.1088/1475-7516/2019/08/011}}.

\bibitem{Saha:2020bis}
P.~Saha, S.~Anand, L.~Sriramkumar, {Accounting for the time evolution of the
  equation of state parameter during reheating} (2020).
\newblock \href {http://arxiv.org/abs/2005.01874} {\path{arXiv:2005.01874}}.

\bibitem{Podolsky:2005bw}
D.~I. Podolsky, G.~N. Felder, L.~Kofman, M.~Peloso, {Equation of state and
  beginning of thermalization after preheating}, Phys. Rev. D73 (2006) 023501.
\newblock \href {http://arxiv.org/abs/hep-ph/0507096}
  {\path{arXiv:hep-ph/0507096}}, \href
  {https://doi.org/10.1103/PhysRevD.73.023501}
  {\path{doi:10.1103/PhysRevD.73.023501}}.

\bibitem{Maity:2018qhi}
D.~Maity, P.~Saha, {(P)reheating after minimal Plateau Inflation and
  constraints from CMB}, JCAP 1907 (2019) 018.
\newblock \href {http://arxiv.org/abs/1811.11173} {\path{arXiv:1811.11173}},
  \href {https://doi.org/10.1088/1475-7516/2019/07/018}
  {\path{doi:10.1088/1475-7516/2019/07/018}}.

\bibitem{Turner:1983he}
M.~S. Turner, {Coherent Scalar Field Oscillations in an Expanding Universe},
  Phys. Rev. D28 (1983) 1243.
\newblock \href {https://doi.org/10.1103/PhysRevD.28.1243}
  {\path{doi:10.1103/PhysRevD.28.1243}}.

\bibitem{Boyanovsky:2003tc}
D.~Boyanovsky, C.~Destri, H.~J. de~Vega, {The Approach to thermalization in the
  classical phi**4 theory in (1+1)-dimensions: Energy cascades and universal
  scaling}, Phys. Rev. D69 (2004) 045003.
\newblock \href {http://arxiv.org/abs/hep-ph/0306124}
  {\path{arXiv:hep-ph/0306124}}, \href
  {https://doi.org/10.1103/PhysRevD.69.045003}
  {\path{doi:10.1103/PhysRevD.69.045003}}.

\bibitem{Dodelson:2003vq}
S.~Dodelson, L.~Hui, {A Horizon ratio bound for inflationary fluctuations},
  Phys. Rev. Lett. 91 (2003) 131301.
\newblock \href {http://arxiv.org/abs/astro-ph/0305113}
  {\path{arXiv:astro-ph/0305113}}, \href
  {https://doi.org/10.1103/PhysRevLett.91.131301}
  {\path{doi:10.1103/PhysRevLett.91.131301}}.

\bibitem{Liddle:2003as}
A.~R. Liddle, S.~M. Leach, {How long before the end of inflation were
  observable perturbations produced?}, Phys. Rev. D68 (2003) 103503.
\newblock \href {http://arxiv.org/abs/astro-ph/0305263}
  {\path{arXiv:astro-ph/0305263}}, \href
  {https://doi.org/10.1103/PhysRevD.68.103503}
  {\path{doi:10.1103/PhysRevD.68.103503}}.

\bibitem{Dufaux:2006ee}
J.~F. Dufaux, G.~N. Felder, L.~Kofman, M.~Peloso, D.~Podolsky, {Preheating with
  trilinear interactions: Tachyonic resonance}, JCAP 0607 (2006) 006.
\newblock \href {http://arxiv.org/abs/hep-ph/0602144}
  {\path{arXiv:hep-ph/0602144}}, \href
  {https://doi.org/10.1088/1475-7516/2006/07/006}
  {\path{doi:10.1088/1475-7516/2006/07/006}}.

\bibitem{Felder:2000hj}
G.~N. Felder, J.~Garcia-Bellido, P.~B. Greene, L.~Kofman, A.~D. Linde,
  I.~Tkachev, {Dynamics of symmetry breaking and tachyonic preheating}, Phys.
  Rev. Lett. 87 (2001) 011601.
\newblock \href {http://arxiv.org/abs/hep-ph/0012142}
  {\path{arXiv:hep-ph/0012142}}, \href
  {https://doi.org/10.1103/PhysRevLett.87.011601}
  {\path{doi:10.1103/PhysRevLett.87.011601}}.

\bibitem{Felder:2001kt}
G.~N. Felder, L.~Kofman, A.~D. Linde, {Tachyonic instability and dynamics of
  spontaneous symmetry breaking}, Phys. Rev. D 64 (2001) 123517.
\newblock \href {http://arxiv.org/abs/hep-th/0106179}
  {\path{arXiv:hep-th/0106179}}, \href
  {https://doi.org/10.1103/PhysRevD.64.123517}
  {\path{doi:10.1103/PhysRevD.64.123517}}.

\bibitem{GarciaBellido:2002aj}
J.~Garcia-Bellido, M.~Garcia~Perez, A.~Gonzalez-Arroyo, {Symmetry breaking and
  false vacuum decay after hybrid inflation}, Phys. Rev. D 67 (2003) 103501.
\newblock \href {http://arxiv.org/abs/hep-ph/0208228}
  {\path{arXiv:hep-ph/0208228}}, \href
  {https://doi.org/10.1103/PhysRevD.67.103501}
  {\path{doi:10.1103/PhysRevD.67.103501}}.

\bibitem{Copeland:2002ku}
E.~J. Copeland, S.~Pascoli, A.~Rajantie, {Dynamics of tachyonic preheating
  after hybrid inflation}, Phys. Rev. D 65 (2002) 103517.
\newblock \href {http://arxiv.org/abs/hep-ph/0202031}
  {\path{arXiv:hep-ph/0202031}}, \href
  {https://doi.org/10.1103/PhysRevD.65.103517}
  {\path{doi:10.1103/PhysRevD.65.103517}}.

\bibitem{Amin:2011hj}
M.~A. Amin, R.~Easther, H.~Finkel, R.~Flauger, M.~P. Hertzberg, {Oscillons
  After Inflation}, Phys. Rev. Lett. 108 (2012) 241302.
\newblock \href {http://arxiv.org/abs/1106.3335} {\path{arXiv:1106.3335}},
  \href {https://doi.org/10.1103/PhysRevLett.108.241302}
  {\path{doi:10.1103/PhysRevLett.108.241302}}.

\bibitem{Antusch:2015nla}
S.~Antusch, D.~Nolde, S.~Orani, {Hill crossing during preheating after hilltop
  inflation}, JCAP 06 (2015) 009.
\newblock \href {http://arxiv.org/abs/1503.06075} {\path{arXiv:1503.06075}},
  \href {https://doi.org/10.1088/1475-7516/2015/06/009}
  {\path{doi:10.1088/1475-7516/2015/06/009}}.

\bibitem{Gleiser:2014ipa}
M.~Gleiser, N.~Graham, {Transition To Order After Hilltop Inflation}, Phys.
  Rev. D 89~(8) (2014) 083502.
\newblock \href {http://arxiv.org/abs/1401.6225} {\path{arXiv:1401.6225}},
  \href {https://doi.org/10.1103/PhysRevD.89.083502}
  {\path{doi:10.1103/PhysRevD.89.083502}}.

\bibitem{DeCross:2015uza}
M.~P. DeCross, D.~I. Kaiser, A.~Prabhu, C.~Prescod-Weinstein, E.~I.
  Sfakianakis, {Preheating after Multifield Inflation with Nonminimal
  Couplings, I: Covariant Formalism and Attractor Behavior}, Phys. Rev. D
  97~(2) (2018) 023526.
\newblock \href {http://arxiv.org/abs/1510.08553} {\path{arXiv:1510.08553}},
  \href {https://doi.org/10.1103/PhysRevD.97.023526}
  {\path{doi:10.1103/PhysRevD.97.023526}}.

\bibitem{DeCross:2016fdz}
M.~P. DeCross, D.~I. Kaiser, A.~Prabhu, C.~Prescod-Weinstein, E.~I.
  Sfakianakis, {Preheating after multifield inflation with nonminimal
  couplings, II: Resonance Structure}, Phys. Rev. D 97~(2) (2018) 023527.
\newblock \href {http://arxiv.org/abs/1610.08868} {\path{arXiv:1610.08868}},
  \href {https://doi.org/10.1103/PhysRevD.97.023527}
  {\path{doi:10.1103/PhysRevD.97.023527}}.

\bibitem{DeCross:2016cbs}
M.~P. DeCross, D.~I. Kaiser, A.~Prabhu, C.~Prescod-Weinstein, E.~I.
  Sfakianakis, {Preheating after multifield inflation with nonminimal
  couplings, III: Dynamical spacetime results}, Phys. Rev. D 97~(2) (2018)
  023528.
\newblock \href {http://arxiv.org/abs/1610.08916} {\path{arXiv:1610.08916}},
  \href {https://doi.org/10.1103/PhysRevD.97.023528}
  {\path{doi:10.1103/PhysRevD.97.023528}}.

\bibitem{Krajewski:2018moi}
T.~Krajewski, K.~Turzyński, M.~Wieczorek, {On preheating in $\alpha$-attractor
  models of inflation}, Eur. Phys. J. C79~(8) (2019) 654.
\newblock \href {http://arxiv.org/abs/1801.01786} {\path{arXiv:1801.01786}},
  \href {https://doi.org/10.1140/epjc/s10052-019-7155-z}
  {\path{doi:10.1140/epjc/s10052-019-7155-z}}.

\bibitem{Iarygina:2018kee}
O.~Iarygina, E.~I. Sfakianakis, D.-G. Wang, A.~Achucarro, {Universality and
  scaling in multi-field $\alpha$-attractor preheating}, JCAP 06 (2019) 027.
\newblock \href {http://arxiv.org/abs/1810.02804} {\path{arXiv:1810.02804}},
  \href {https://doi.org/10.1088/1475-7516/2019/06/027}
  {\path{doi:10.1088/1475-7516/2019/06/027}}.

\bibitem{Nguyen:2019kbm}
R.~Nguyen, J.~van~de Vis, E.~I. Sfakianakis, J.~T. Giblin, D.~I. Kaiser,
  {Nonlinear Dynamics of Preheating after Multifield Inflation with Nonminimal
  Couplings}, Phys. Rev. Lett. 123~(17) (2019) 171301.
\newblock \href {http://arxiv.org/abs/1905.12562} {\path{arXiv:1905.12562}},
  \href {https://doi.org/10.1103/PhysRevLett.123.171301}
  {\path{doi:10.1103/PhysRevLett.123.171301}}.

\bibitem{vandeVis:2020qcp}
J.~van~de Vis, R.~Nguyen, E.~I. Sfakianakis, J.~T. Giblin, D.~I. Kaiser,
  {Time-Scales for Nonlinear Processes in Preheating after Multifield Inflation
  with Nonminimal Couplings} (5 2020).
\newblock \href {http://arxiv.org/abs/2005.00433} {\path{arXiv:2005.00433}}.

\bibitem{Martin:2020fgl}
J.~Martin, T.~Papanikolaou, L.~Pinol, V.~Vennin, {Metric preheating and
  radiative decay in single-field inflation}, JCAP 05 (2020) 003.
\newblock \href {http://arxiv.org/abs/2002.01820} {\path{arXiv:2002.01820}},
  \href {https://doi.org/10.1088/1475-7516/2020/05/003}
  {\path{doi:10.1088/1475-7516/2020/05/003}}.

\bibitem{Felder:2007nz}
G.~N. Felder, {CLUSTEREASY: A program for lattice simulations of scalar fields
  in an expanding universe on parallel computing clusters}, Comput. Phys.
  Commun. 179 (2008) 604--606.
\newblock \href {http://arxiv.org/abs/0712.0813} {\path{arXiv:0712.0813}},
  \href {https://doi.org/10.1016/j.cpc.2008.06.002}
  {\path{doi:10.1016/j.cpc.2008.06.002}}.

\bibitem{Felder:2000hq}
G.~N. Felder, I.~Tkachev, {LATTICEEASY: A Program for lattice simulations of
  scalar fields in an expanding universe}, Comput. Phys. Commun. 178 (2008)
  929--932.
\newblock \href {http://arxiv.org/abs/hep-ph/0011159}
  {\path{arXiv:hep-ph/0011159}}, \href
  {https://doi.org/10.1016/j.cpc.2008.02.009}
  {\path{doi:10.1016/j.cpc.2008.02.009}}.

\bibitem{Figueroa:2020rrl}
D.~G. Figueroa, A.~Florio, F.~Torrenti, W.~Valkenburg, {The art of simulating
  the early Universe -- Part I} (6 2020).
\newblock \href {http://arxiv.org/abs/2006.15122} {\path{arXiv:2006.15122}}.

\end{thebibliography}
 \bibliographystyle{elsarticle-num} 
 
\end{document}